\documentclass[12pt]{article}
\usepackage{epsfig}

\catcode`\@=11
\DeclareMathAlphabet{\mathpzc}{OT1}{pzc}{m}{it}

\newcommand{\insertfig}[2]{\mbox{\epsfxsize=#1cm \epsfbox{#2.eps}}}

\newcommand{\cQ}{{\cal Q}}
\newcommand{\Bx}{x_{\rm B}}
\newcommand{\GeV}{{\rm GeV}}
\font\cmss=cmss12 
\def\1{\hbox{{1}\kern-.25em\hbox{l}}}
\def\bfZ{\relax{\hbox{\cmss Z\kern-.4em Z}}}

\def \be  {\begin{equation}}
\def \ee  {\end{equation}}
\def \ba  {\begin{eqnarray}}
\def \ea  {\end{eqnarray}}
\def \baa {\begin{eqnarray*}}
\def \eaa {\end{eqnarray*}}
\def \bb  {\begin {thebibliography} }
\def \eb  {\end{thebibliography}}
\def \lab #1 {\label{#1}}

\newcommand\re[1]{(\ref{#1})}

\def \matrix #1 {\left(\begin{array}{cc} #1 \end{array}\right)}

\newcommand{\as}{\ifmmode\alpha_{\rm s}\else{$\alpha_{\rm s}$}\fi}
\newcommand{\asbar}{\ifmmode\bar{\alpha}_{\rm s}\else{$\bar{\alpha}_{\rm s}$}\fi}

\newcommand{\ft}[2]{{\textstyle\frac{#1}{#2}}}

\font\cmss=cmss12 
\def\inbar{\,\vrule height1.5ex width.4pt depth0pt}
\def\IC{\relax\hbox{$\inbar\kern-.3em{\rm C}$}}
\def\IZ{\relax{\hbox{\cmss Z\kern-.4em Z}}}
\def\IR{{\hbox{{\rm I}\kern-.2em\hbox{\rm R}}}}

\def\IP{{\hbox{{\rm I}\kern-.2em\hbox{\rm P}}}}
\def\II{\hbox{{1}\kern-.25em\hbox{l}}}

\def\numberbysection{\@addtoreset{equation}{section}
                     \def\theequation{\thesection.\arabic{equation}}}
\numberbysection

\newbox\lett\newdimen\lheight\newdimen\lwidth
\def\ontop#1#2{\setbox\lett=\hbox{#2}\lheight\ht\lett
\multiply\lheight by 12 \divide\lheight by 10\relax%
\lwidth\wd\lett \multiply\lwidth by 8 \divide\lwidth by 10\relax #2\kern-\lwidth%
\raise\lheight\hbox{{$\scriptstyle #1$}}\kern.1ex}

\def\XXint#1#2#3{{\setbox0=\hbox{$#1{#2#3}{\int}$}
     \vcenter{\hbox{$#2#3$}}\kern-.5\wd0}}

\begin{document}

\begin{titlepage}

\thispagestyle{empty}

\vskip2cm

\centerline{\large \bf Refined analysis of photon leptoproduction off spinless target}

\vspace{15mm}

\centerline{\sc A.V. Belitsky$^a$, D. M\"uller$^b$}

\vspace{15mm}

\centerline{\it $^a$Department of Physics, Arizona State University}
\centerline{\it Tempe, AZ 85287-1504, USA}

\vspace{5mm}

\centerline{\it $^b$Institut f\"ur Theoretische Physik II, Ruhr-Universit¨at Bochum}
\centerline{\it D-44780 Bochum, Germany}

\vspace{3cm}

\centerline{\bf Abstract}

\vspace{5mm}

We calculate the differential cross section for real photon electroproduction off spinless
hadron which sevres as a main probe of the hadrons structure via the concept of generalized
parton distributions. Compared to previously available computations performed with twist-three
power accuracy, we exactly accounted for all kinematical effects in hadron mass and momentum
transfer which arise from leptonic helicity amplitudes. We performed numerical studies of these
kinematical effects and demonstrated that in the valence quark region and rather low virtualities
of the hard photon which sets the factorization scale, the available approximate results
significantly overestimate the cross section rates in comparison to exact formulas.

\end{titlepage}

\setcounter{footnote} 0

\newpage

\pagestyle{plain}
\setcounter{page} 1

\section{Introduction}

The qualitative mechanism which binds hadronic constituents together in a bound state is well
understood within the framework of QCD via the formation of collimated gluon flux tubes between
quarks. This picture is confirmed by lattice gauge theory simulations which aim at a quantitative
exploration of the hadronic structure. However, physical observables which can be measured in
high-energy experiments necessarily involve correlation functions of elementary quark and gluon
fields separated by light-like distances and thus evade straightforward use of euclidean lattice
tools. Therefore, in the lack of other analytical/numerical first-principle techniques,
phenomenological analyses of experimental data are currently the only viable alternative route to
unravel manifestations of the intricate bound state problem in QCD.

Among hadronic observables, generalized parton distributions (GPDs) \cite{MueRobGeyDitHor94} are
the most elaborate light-cone correlation functions containing simultaneous information on both
position and momentum distributions of strongly interacting constituents \cite{ImpPar00,BelJiYua04}.
Similarly to conventional collinear parton densities which are measured in deeply inelastic
scattering, GPDs can be probed in experiments involving electroweak bosons with production of a
real photon or mesons in the final state (see Ref. \cite{Rev04} for reviews). While the meson
production receives contamination from additional hadronic final states, the former are free from
these uncertainties and provide a clean access to hadronic inner content through GPDs.

To date, the most complete analytic calculations were performed for photon leptoproduction cross section
off spin-zero and spin-one-half targets and were limited to twist-three accuracy. This approximation
implies that only terms suppressed by a single power of the hard photon virtuality were kept in all
analytical expressions, corresponding both to leptonic and hadronic parts of amplitudes. The latter
is defined by the expectation value of the chronological product of two electromagnetic currents
between in and out hadronic states,
\be
\label{DVCSamplitude}
T_{\mu\nu} = i \int d^4 z \, {\rm e}^{\frac{i}{2} (q_1 + q_2) \cdot z}
\langle p_2 | T \{ j_\mu (z/2) j_\nu (- z/2) \} | p_1 \rangle
\, .
\ee
This tensor is parameterized in terms of the so-called Compton form factors (CFFs) $\mathcal{F}
(\xi, t; \mathcal{Q})$ which enter as coefficients in front of independent Lorentz structures.
The CFFs depend on the generalized Bjorken-like scaling variable $\xi$, the squared momentum
transfer $t$, and the photon virtuality $q_1^2 = - {\cal Q}^2$. The QCD factorization theorems
are indispensable in separating CFFs in terms of short-distance coefficient functions
$C (x, \xi; \mathcal{Q}/\mu)$, controllable via conventional perturbation theory in QCD coupling
constant, and long-distance dynamics encoded in GPDs $F (x, \xi, t; \mu)$,
\be
\label{DefCFFs}
\mathcal{F} (\xi, t; \mathcal{Q})
=
\int dx \, C (x, \xi; \mathcal{Q}/\mu) \, F (x, \xi, t; \mu)
\, .
\ee
The state-of-the-art considerations of the hadronic tensor \re{DVCSamplitude} were done in the
twist-three approximation \cite{AniPirTer00,PenPolShuStr00,BelMue00a,RadWei00,KivPolSchTer00}. The
hierarchy of hadronic matrix elements of higher twist operators emerging in the operator product
expansion of Eq.\ \re{DVCSamplitude} suggests smallness of their effect on event rates even at rather
low virtualities. This phenomenon is well-known in deeply inelastic scattering and yields precocious
scaling of corresponding observables making the neglect of operators of twist-four and higher
legitimate. On the other hand, the approximation to merely leading and first subleading contributions
stemming from the leptonic tensor were an artifact of matching the expansion of hadronic and leptonic
parts. When the latter approximation is waived, numerical considerations demonstrate significant
deviations between the two predictions for the kinematics of Jefferson Lab experiments. Therefore, in
the present paper we perform a refined analysis starting with leptoproduction cross section of real
photons off a spinless target.

Our subsequent consideration is organized as follows. In the next section, we present a brief
profile of the twist-three formalism we have developed in our earlier work \cite{BelMueKir01}.
In Section \ref{HelicityAmplitudes}, we introduce the formalism of helicity amplitudes
\cite{KroSchGui96,DieGouPirRal97} which proves to be very efficient in separating power suppressed
effects arising from strong-coupling dynamics in the form of higher-twist correlations, on the one
hand, and kinematical effects due to nonvanishing masses of hadrons and momentum transfer in the
$t-$channel, on the other. As previously, the hadronic part is calculated to twist-three accuracy,
while we account for aforementioned kinematically-suppressed contributions exactly and thus keep
all power effects in the leptonic part. As a consequence, the very transparent classification scheme
of Ref.\ \cite{BelMueKir01}, which allows one to identify Fourier harmonics in the azimuthal angle
with specific twists of contributing GPDs, ceases its existence at small photon virtuality in the
valence region. Then in Sections \ref{SectionLeptonicTensor} and \ref{SectionHadronicTensor}, we
provide an estimate of different contributions to the cross section to get a handle on the most
sizable effects. Finally, we conclude. The discussion of the kinematics is deferred to Appendix
\ref{KinematAppend}, while the explicit expressions for leptonic helicity amplitudes are summarized
in Appendix \ref{FourierHarmonics}.

\section{Electroproduction cross section}
\label{Sec-AziAngDep}

\begin{figure}[t]
\vspace{-2.5cm}
\begin{center}
\mbox{
\begin{picture}(0,173)(275,0)
\put(70,0){\insertfig{14}{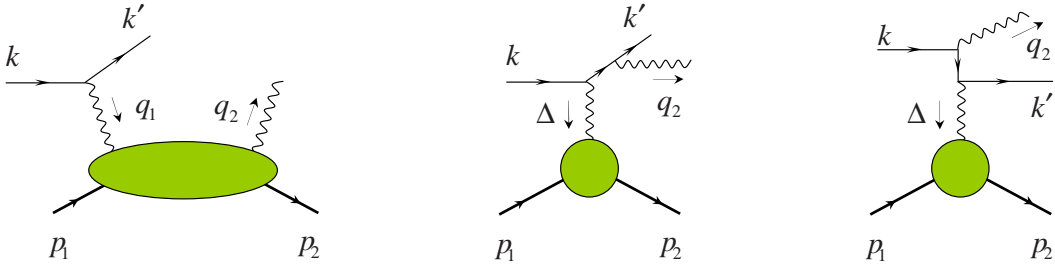}}
\end{picture}
}
\end{center}
\caption{\label{DVCSandBHamplitudes} Amplitudes contributing to the photon leptoproduction
cross section. The first one (i.e, the left-most) is the DVCS amplitude factorized into GPDs
while the other two are the Bethe-Heitler amplitudes parameterized by hadronic electromagnetic
form factors.}
\end{figure}

The main focus of our present analysis is the four-fold cross section for scattering of a light
lepton $\ell = e^\mp$ off a spinless hadron $h$ and production of a photon in the final state,
$\ell (k) h (p_1) \to \ell (k^\prime) h (p_2) \gamma (q_2)$,
\be
\label{WQ}
d \sigma
=
\frac{\alpha^3  \Bx y } {8 \, \pi \,  {\cal Q}^2 \sqrt{1 + \epsilon^2}}
\left| \frac{\cal T}{e^3} \right|^2
d \Bx dy d |t| d \phi
\, .
\ee
The phase space of the process is parameterized by the Bjorken variable $\Bx = \mathcal{Q}^2/(2 p_1
\cdot q_1)$ defined by the virtual photon euclidean mass $\mathcal{Q}^2 = - q_1^2$ of momentum
$q_1 = k - k'$, the squared momentum transfer $t \equiv \Delta^2$ with $\Delta = p_2 - p_1$, the
lepton energy loss $y = p_1\cdot q_1/p_1\cdot k$, and finally the azimuthal angle $\phi$ of the
outgoing hadron. The dependence on the latter provides a very important handle on different
combinations of twist-two and -three GPDs which enter the hadronic amplitudes \re{DVCSamplitude}.
The cross section depends on small kinematical parameters which we will account for exactly in the
present consideration. One of them has already appeared explicitly in Eq.\ \re{WQ} and is given by
the ratio of the hadronic mass $M$ to the photon virtuality $\mathcal{Q}$,
\be
\epsilon \equiv 2 \Bx \frac{M}{{\cal Q}}
\, .
\ee
The other will be introduced below.

According to Fig.\ \ref{DVCSandBHamplitudes}, the amplitude of the process ${\cal T}$ is a sum
of two distinct contributions, one involving the deeply virtual Compton scattering (DVCS)
tensor \re{DVCSamplitude} and termed ${\cal T}^{\rm DVCS}$ and the other one with leptonic
Bethe-Heitler (BH) subprocess coupled to the hadronic electromagnetic current $J_\mu$
parameterized via the (pseudo)scalar form factor $F (t)$ as
\be
J_\mu = \langle p_2 | j_\mu (0) | p_1 \rangle = (p_1 + p_2)_\mu F (t)
\, ,
\ee
and dubbed ${\cal T}^{\rm BH}$. The latter is real (to the lowest order in the QED fine structure
constant) and $F (t)$ is taken from other measurements. The azimuthal angular dependence of each
of the three terms in
\begin{equation}
{\cal T}^2
= |{\cal T}^{\rm BH}|^2 + |{\cal T}^{\rm DVCS}|^2 + {\cal I}
\, ,
\end{equation}
with the interference term
\begin{equation}
{\cal I}
= {\cal T}^{\rm DVCS} ( {\cal T}^{\rm BH} )^\ast
+ ( {\cal T}^{\rm DVCS} )^\ast {\cal T}^{\rm BH}
\, ,
\end{equation}
arises from the Lorentz-invariant scalar products defining the leptonic and hadronic parts of
amplitudes as explained at length below. Since the square of the Bethe-Heitler amplitude was
computed exactly in Ref.\ \cite{BelMueKirSch00}, we will focus our attention on the remaining
contributions. Both of them are expressed by contractions of leptonic tensors with corresponding
hadronic transition amplitudes, involving either the square of the DVCS amplitudes for
$|{\cal T}^{\rm DVCS}|^2$, or being linear both in DVCS and hadronic electromagnetic current
for $\mathcal{I}$,
\begin{eqnarray}
\label{Def-Sqa-DVCS}
&&
|{\cal T}^{\rm DVCS}|^2
=
- \frac{e^6}{{\cal Q}^2} L^{\mu\nu}(\lambda) T_{\tau \mu} (T_{\phantom{\tau }\nu}^{\tau})^\ast
\, , \\
\label{Def-Sqa-Int}
&&
{\cal I}
=
\frac{\pm e^6}{\Delta^2 {\cal P}_1(\phi) {\cal P}_2(\phi)}
\left\{
L_{\mu\nu\tau}(\lambda)  T^{\tau\mu} (J^{\nu})^\ast
+
L^\ast _{\mu\nu\tau}(\lambda)   (T^{\tau\mu})^\ast  J^{\nu}\right\}
\, .
\end{eqnarray}
Here the $+$ ($-$) sign in the interference stands for the negatively (positively) charged lepton
beam. The leptonic tensors to the lowest order in the fine structure constant read\footnote{We adopt
the conventions for Dirac matrices and Lorentz tensors from Itzykson and Zuber \cite{ItzZub80},
e.g., $\varepsilon^{0123} = + 1$. We assume that the lepton helicity is positive, i.e., $\lambda
= +1$ if the spin is aligned with the direction of the lepton three-momentum.}, respectively,
\begin{eqnarray}
\label{Def-LepTen-DVCS}
L_{\mu\nu} (\lambda)
\!\!\!&=&\!\!\!
\frac{2}{{\cal Q}^2}
\left(
k_\mu k^\prime_\nu + k_\nu k^\prime_\mu-k\cdot k^\prime g_{\mu\nu}
+
i \lambda  \varepsilon_{\mu\nu k k^\prime}
\right) , \\
\label{Def-LepTen-I}
L_{\mu\nu\tau}(\lambda)
\!\!\!&=&\!\!\!
\frac{(k - q_2)^2 (k - \Delta)^2}{{\cal Q}^6}
{\rm tr}\, \ft12 (1 - \lambda \gamma_5)
\left\{ \gamma_\nu \left( {\not\!k} - {\not\!\!\Delta} \right)^{-1}
\gamma_\tau + \gamma_\tau \left( {\not\!k}^\prime + {\not\!\!\Delta} \right)^{-1}
\gamma_\nu {\not\!k}^\prime
\right\} \gamma_\mu \, {\not\!k}
\, , \qquad
\end{eqnarray}
where the lepton mass has been set to zero. The rescaled BH propagators
\begin{equation}
\label{ExaBHpro}
{\cal P}_1 \equiv \frac{(k - q_2)^2}{{\cal Q}^2} = 1 + \frac{2k\cdot \Delta}{{\cal Q}^2}
\, , \qquad
{\cal P}_2 \equiv \frac{(k - \Delta)^2}{{\cal Q}^2} = \frac{t - 2 k \cdot \Delta}{{\cal Q}^2}
\, ,
\end{equation}
emerge as contaminating sources of the azimuthal angle dependence which interfere with Fourier
harmonics accompanying the generalized Compton form factors if expanded in inverse powers of
the large photon virtuality $\mathcal{Q}^2$. Thus they will be treated exactly.

\begin{figure}[t]
\vspace{-2.5cm}
\begin{center}
\mbox{
\begin{picture}(0,320)(250,0)
\put(70,0){\insertfig{11}{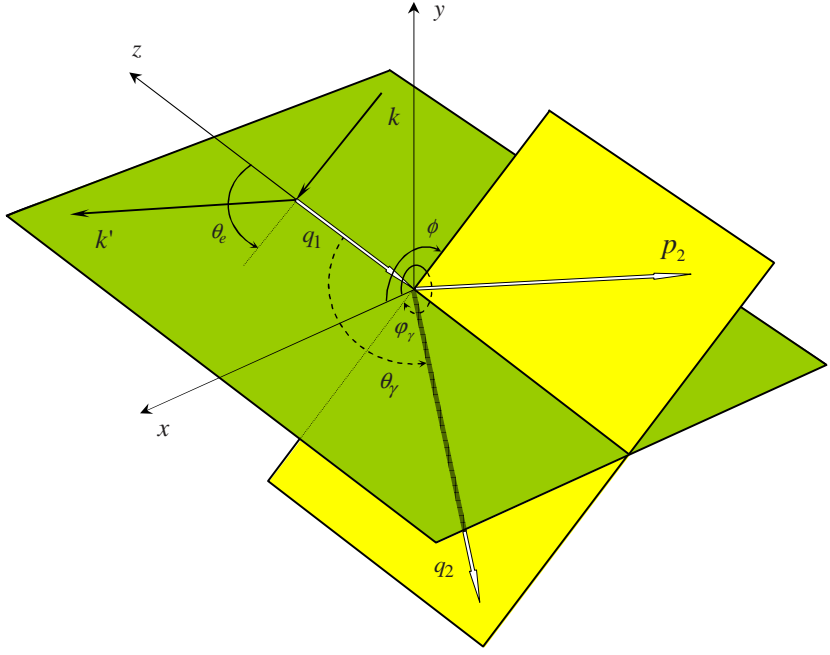}}
\end{picture}
}
\end{center}
\caption{\label{Fig-Kin} The kinematics of the leptoproduction in the target rest frame. The
$z$-direction is chosen counter-along the three-momentum of the incoming virtual photon. The
lepton three-momenta form the lepton scattering plane, while the recoiled proton and outgoing
real photon define the hadronic scattering plane. In this reference system the azimuthal angle
of the scattered lepton vanishes, while the azimuthal angle between the lepton plane and the
recoiled proton momentum is $\phi$.}
\end{figure}

To continue our discussion we choose the target rest frame as shown in Fig.\ \ref{Fig-Kin}.
The explicit components of particle's momenta in this frame are defined in Appendix
\ref{KinematAppend}. Guarded with these, one immediately computes all invariant products in
terms of variables of the phase space of the process. For instance, one finds for the
angular dependence of the BH propagators,
\begin{equation}
\label{kDelta}
k \cdot \Delta
= - \frac{{\cal Q}^2}{2y (1 + \epsilon^2)}
\Bigg\{
1 + 2 K \cos{\phi} - \frac{t}{{\cal Q}^2}
\left( 1 - \Bx (2-y) + \frac{y \epsilon^2}{2} \right)
+  \frac{y \epsilon^2}{2}
\Bigg\} \, ,
\end{equation}
where the $1/{\cal Q}$-power suppressed kinematical $K-$factor, also showing up below in the
Fourier expansion (\ref{AmplitudesSquared}) -- (\ref{InterferenceTerm}), reads
\begin{equation}
\label{Kfactor}
K^2 = -\frac{t^\prime}{{\cal Q}^2} (1 - \Bx)
\left( 1 - y - \frac{y^2\epsilon^2}{4} \right)
\left\{
\sqrt{1 + \epsilon^2}
+ \frac{4\Bx (1 - \Bx) + \epsilon^2}{4(1 - \Bx)}
\frac{t^\prime}{{\cal Q}^2}
\right\} \, ,
\end{equation}
with the plus sign taken for the square root in Eq.\ (\ref{kDelta}) and the variable $t^\prime$
standing for
\be
t^\prime = t - t_{\rm min}
\, .
\ee
The variable $K$ vanishes at the kinematical boundary $t = t_{\rm min}$,
determined by the minimal value of the momentum transfer in $t-$channel
\begin{eqnarray}
\label{Def-tmin}
t_{\rm min}
= - {\cal Q}^2
\frac{2(1 - \Bx) \left(1 - \sqrt{1 + \epsilon^2}\right) + \epsilon^2}
{4\Bx (1 - \Bx) + \epsilon^2}
\, ,
\end{eqnarray}
as well as at maximal value of the lepton energy loss $y_{\rm max} = 2 \epsilon^{- 2} (\sqrt{1 +
\epsilon^2} - 1)$.

\section{BKM approximation}
\label{Sec-BKM}

In the frame we have chosen for our analysis, the contractions between the leptonic and hadronic
tensor structures yield finite sums of Fourier harmonics, whose maximal frequencies are defined by
the the rank-$m$ of the leptonic tensor in the incoming lepton momentum $k_\mu$. Note, however,
that the polarization-dependent part of the leptonic tensors possesses one power of the four-vector
$k_\mu$ less than in the unpolarized sector. As a consequence, the highest harmonic accompanied by
the lepton helicity $\lambda$ will be $\sin([m-1]\phi)$ rather than $\sin(m\phi)$, such that one
finds for squared amplitudes
\begin{eqnarray}
\label{AmplitudesSquared}
&& |{\cal T}^{\rm DVCS}|^2
=
\frac{e^6}{y^2 {\cal Q}^2}\left\{
c^{\rm DVCS}_0
+ \sum_{n=1}^2
\left[
c^{\rm DVCS}_n \cos (n\phi) + s^{\rm DVCS}_n \sin (n \phi)
\right]
\right\} \, ,
\\
\label{InterferenceTerm}
&&{\cal I}
= \frac{\pm e^6}{\Bx y^3 t {\cal P}_1 (\phi) {\cal P}_2 (\phi)}
\left\{
c_0^{\cal I}
+ \sum_{n = 1}^3
\left[
c_n^{\cal I} \cos(n \phi) +  s_n^{\cal I} \sin(n \phi)
\right]
\right\} \, .
\end{eqnarray}
The generation of new harmonics in the azimuthal angle terminates at the twist-three level.
The leading term (in inverse powers of the hard scale $\mathcal{Q}$) in each Fourier
coefficient\footnote{We adopt here the notation of Ref.~\cite{BelMueKir01} rather than of
Ref.\ \cite{BelMueKirSch00}.} was computed in Ref.\ \cite{BelMueKirSch00} and will be presented
below for the sake of comparison with improved approximations computed in the next section. Here
we will point out that the leading contributions to $c^{\cal I}_1, s^{\cal I}_1$ as well as
$c^{\rm DVCS}_0$ emerge from twist-two GPDs \cite{DieGouPirRal97,BelMueNieSch00}. The rest of
the Fourier harmonics provides an additional angular dependence and is given in terms of twist-two,
i.e., for $c^{\cal I}_0$, and twist-three GPDs, i.e., for $c^{\rm DVCS}_1$, $s^{\rm DVCS}_1$,
$c^{\cal I}_2$, and $s^{\cal I}_2$. The harmonics proportional to $\cos{(3\phi)}$ [$\cos{(2\phi)}$]
or $\sin{(3\phi)}$ [$\sin{(2\phi)}$] in the interference [squared DVCS] term stem from the
twist-two double helicity-flip gluonic GPDs alone. They are not contaminated by any twist-two quark
amplitudes, however, will be affected by twist-four power corrections \cite{KivMan01}. We neglect
in our consequent considerations the effects of dynamical higher-twist (larger than three)
contributions by the token alluded to in the introduction.

\subsection{Squared DVCS amplitude}
\label{SubSec-AziAngDep-DVCS}

The Fourier coefficients of $|{\cal T}^{\rm DVCS}|^2$ naturally split into the product of factors
depending on the leptonic kinematical variables and hadronic functions ${\cal C}^{\rm DVCS}$
\begin{eqnarray}
\label{Res-Mom-DVCS-0}
c^{\rm DVCS}_{0}
\!\!\!&=&\!\!\!
2 ( 2 - 2 y + y^2 )
{\cal C}^{\rm DVCS}
\left(
{\cal H},{\cal H}^\ast; {\cal H}_T,{\cal H}_T^\ast
\right)  ,
\\
\label{Res-Mom-DVCS-1}
\left\{
{c^{\rm DVCS}_{1} \atop s^{\rm DVCS}_{1}}
\right\}
\!\!\!&=&\!\!\!
\frac{8 K}{2 - \Bx} \left\{ { (2-y) \, \Re{\rm e} \atop - \lambda y \, \Im{\rm m} } \right\}
\, {\cal C}^{\rm DVCS}
\left(
{\cal H}^{\rm eff};{\cal H}^\ast; {\cal H}_T^\ast\right) \, ,
\\
\label{Res-Mom-DVCS-2}
c^{\rm DVCS}_{2}
\!\!\!&=&\!\!\!
\frac{16 {\cal Q}^2 K^2}{M^2 (2 - \Bx)^2}
\Re{\rm e}\,
{\cal C}^{\rm DVCS}_{T} \left({\cal H}, {\cal H}_T^\ast\right)
\, .
\end{eqnarray}
The latter are bilinear in the CFFs and respectively read
\begin{eqnarray}
\label{Def-C-DVCS-unp}
{\cal C}^{\rm DVCS}
\left(
{\cal H},{\cal H}^\ast,{\cal H}_T, {\cal H}_T^\ast
\right)
\!\!\!&=&\!\!\! {\cal H} {\cal H}^\ast + \frac{{\widetilde K}^4}{(2-\Bx)^4}
{\cal H}_T {\cal H}_T^\ast \,,
\nonumber\\
{\cal C}^{\rm DVCS} \left(
{\cal H}_{\rm eff};{\cal H}^\ast;{\cal H}_T^\ast
\right) \!\!\!&=&\!\!\!
{\cal H}_{\rm eff}
\left({\cal H}^\ast + \frac{{2 \widetilde K}^2}{M^2 (2-\Bx)^2}{\cal H}_T^\ast \right)
\\
{\cal C}^{\rm DVCS}_{T} \left(
{\cal H},{\cal H}_T^\ast
\right) \!\!\!&=&\!\!\! {\cal H} {\cal H}_T^\ast
\, .
\nonumber
\end{eqnarray}
At this point, we would like to recall that the spinless hadron acquires only two types of
leading twist GPDs, unpolarized quark GPD $H$ and gluon transversity GPD $H_T$ giving
corresponding names to the CFFs \re{DefCFFs}. In the adopted approximation, the
$c^{\rm DVCS}_{0}$ harmonic is expressed via the twist-two CFF ${\cal H}$, while the
coefficients $c^{\rm DVCS}_{1}$ and $s^{\rm DVCS}_{1}$ arise from the interference of
twist-two and effective twist-three CFFs,
\begin{eqnarray}
\label{Tw3Eff}
{\cal H}^{\rm eff} \equiv - 2 \xi
\left(
\frac{1}{1 + \xi} {\cal H} + {\cal H}^3_+ - {\cal H}^3_-
\right) \, ,
\end{eqnarray}
with the CFFs $\mathcal{H}^3_\pm$ related to functions $H^3_\pm$ given by a convolution of the
twist-two GPD $H$ and the so-called Wandzura-Wilczek kernel provided one neglects dynamical
quark-gluon-quark correlation functions \cite{BelMueKirSch00}. Here the generalized Bjorken
variable $\xi$ is related to the usual one $\Bx$ via $\xi \simeq \Bx/(2 - \Bx)$. Finally, the
Fourier coefficients $c^{\rm DVCS}_{2}$ and $s^{\rm DVCS}_{2}$ are induced by the gluon transversity
CFFs\footnote{These were omitted in our earlier consideration \cite{BelMueKirSch00} of DVCS on
spinless targets since they enter the amplitudes suppressed by a power of the QCD coupling
constant. Note that ${\cal H}_T$ also enters as an $(\alpha_s/\pi)^2$ suppressed contribution
to $c^{\rm DVCS}_0$ and  as a $\alpha_s/\pi$ suppressed effect in the twist-three harmonics
$c^{\rm DVCS}_1$ and  $s^{\rm DVCS}_1$, c.f.\ Eq.\ \re{Def-C-DVCS-unp}.}.

\subsection{Interference of Bethe-Heitler and DVCS amplitudes}
\label{SubSec-AziAngDep-INT}

Next, let us present approximate results for the interference term which is the most promising
observable for the phenomenology of GPDs since it is linear in CFFs. This property simplifies
the disentanglement of CFFs from experimental measurements. The Fourier harmonics have the
form:
\begin{eqnarray}
\label{Res-IntTer-unp-c0}
c^{\cal I}_{0}
\!\!\!&=&\!\!\!
- 8  (2 - y)
\Re{\rm e}
\Bigg\{
\frac{(2 - y)^2}{1-y} K^2
+
\frac{t}{{\cal Q}^2} (1 - y)(2 - \Bx)
\Bigg\} {\cal C}^{\cal I} \left({\cal H}\right)\, ,
\\
\label{Res-IntTer-unp}
\left\{{c^{\cal I}_{1} \atop s^{\cal I}_{1}}\right\}
\!\!\!&=&\!\!\!
8 K
\left\{  {-(2 - 2y + y^2) \, \Re{\rm e} \atop \lambda y (2-y) \, \Im{\rm m}} \right\}
{\cal C}^{\cal I}\left({\cal H} \right) \, ,
\\
\label{Res-IntTer-unp-tw3}
\left\{ c^{\cal I}_{2} \atop  s^{\cal I}_{2} \right\}
\!\!\!&=&\!\!\!
\frac{16 K^2}{2 - \Bx}  \left\{ { -(2 - y) \, \Re{\rm e} \atop \lambda y \, \Im{\rm m}} \right\}
{\cal C}^{\cal I}
\left({\cal H}^{\rm eff}\right) \, ,
\\
\label{Res-IntTer-Tra-unp}
c^{\cal I}_{3}
\!\!\!&=&\!\!\! - \frac{8 {\cal Q}^2 K^3}{M^2 (2 - \Bx)^2}
\Re{\rm e}\, {\cal C}^{\cal I}_{T}
\left( {\cal H}_T \right) \, .
\end{eqnarray}
Explicit calculations demonstrate that the twist-three harmonics, i.e., $c^{\cal I}_{2}$
and $s^{\cal I}_{2}$ have the same functional dependence as the twist-two coefficients,
$c^{\cal I}_{1}$ and $s^{\cal I}_{1}$. However, this is not the case for $ c^{\cal I}_{3}$
and $s^{\cal I}_{3}$ which emerge due to gluon helicity-flip CFF ${\cal H}_T$,
\begin{eqnarray}
\label{Def-C-Int-unp}
{\cal C}^{\cal I} ( {\cal H})
\!\!\!&=&\!\!\!
F  {\cal H}
\, , \qquad
{\cal C}^{\cal I}_{T}
=
F {\cal H}_T\, .
\end{eqnarray}
This set of formulas forms the complete result for the real-photon leptoproduction cross
section in the twist-three approximation.

\section{Helicity amplitudes}
\label{HelicityAmplitudes}

The consideration of the previous section was restricted to the twist-three approximation
for the dynamical as well as kinematical effects. Higher-twist operator product expansion
analyses of the off-forward Compton scattering amplitude, similar to the one performed for the
deeply inelastic scattering \cite{EllFurPet83}, is intrinsically involved due to complications
and ambiguities in the choice of operator basis. The incorporation of kinematical power-suppressed
effects is relatively straightforward. In order to achieve this in the most efficient manner we
separate power corrections that arise from the leptonic and hadronic parts by evaluating photon
helicity amplitudes utilizing the polarization vectors for the incoming and outgoing photons in
the target rest frame as defined in Appendix \ref{KinematAppend} in Eqs.\ \re{Polarization1} and
\re{Polarization2}, respectively.

\subsection{Squared DVCS term}

Using the completeness relations \re{Rel-ComHel-1} and \re{Rel-ComHel-2} for the photon
polarization vectors, we can rewrite the square of the DVCS amplitude \re{Def-Sqa-DVCS}
as follows
\begin{eqnarray}
|{\cal T}^{\rm DVCS}|^2
=
\frac{1}{{\cal Q}^2}
\sum_{a = {\scriptscriptstyle -}, 0, {\scriptscriptstyle +}}
\sum_{b = {\scriptscriptstyle -}, 0, {\scriptscriptstyle +}}
{\cal L}_{ab} (\lambda, \phi) {\cal W}_{ab}
\, , \quad
{\cal W}_{ab}
=
{\cal T}^{\rm DVCS}_{a +} \left({\cal T}^{\rm DVCS}_{b +}\right)^\ast
+
{\cal T}^{\rm DVCS}_{a -} \left({\cal T}^{\rm DVCS}_{b -}\right)^\ast
\, ,
\end{eqnarray}
in terms of helicity amplitudes labelled by the helicity states of the (initial) photon. These
are defined by contractions of the Lorentz covariant amplitudes with the photon polarization
vectors \re{Polarization1} and \re{Polarization2},
\be
{\cal L}_{ab}(\lambda,\phi) =
\varepsilon^{\mu\ast}_1(a){\cal L}_{\mu \nu}(\lambda)\varepsilon^{\nu}_1(b)\,,
\ee
and
\be
\label{DVCS2helicity}
{\cal T}^{\rm DVCS}_{ac}(\phi) = (-1)^{a-1}
\varepsilon^{\mu\ast}_2(c) T_{\mu \nu} \varepsilon^{\nu}_1(a) \, ,
\ee
where the phase $(-1)^{a-1}$ takes care of the signature factor in the completeness
relation (\ref{Rel-ComHel-1}).

The helicity amplitudes \re{DVCS2helicity} are constrained by the parity conservation and, as
a consequence, we have just three independent functions,
\begin{eqnarray}
{\cal T}^{\rm DVCS}_{++} =  {\cal T}^{\rm DVCS}_{--}
\, , \quad
{\cal T}^{\rm DVCS}_{0+} =  {\cal T}^{\rm DVCS}_{0-}
\, , \quad
{\cal T}^{\rm DVCS}_{-+} =  {\cal T}^{\rm DVCS}_{+-}
\, .
\end{eqnarray}
The leptonic helicity amplitudes are calculated exactly from the definitions \re{Def-LepTen-DVCS}
and \re{Polarization1} and read
\begin{eqnarray}
{\cal L}_{++} (\lambda)
\!\!\!&=&\!\!\!
\frac{1}{ y^2(1 + \epsilon^2)} \left(2 - 2 y + y^2 + \frac{\epsilon ^2}{2} y^2 \right)
-
\frac{2-y}{\sqrt{1+\epsilon^2} y} \lambda
\, , \\
{\cal L}_{00}
\!\!\!&=&\!\!\!
\frac{4}{y^2(1+\epsilon^2)} \left(1-y + \frac{\epsilon ^2}{4} y^2\right)
\, , \\
{\cal L}_{0+}(\lambda,\phi)
\!\!\!&=&\!\!\!
\frac{2 - y - \lambda y \sqrt{1 + \epsilon^2}}{y^2 (1 + \epsilon^2)}
\sqrt{2} \sqrt{1 - y - \frac{\epsilon ^2}{4} y^2} \, e^{-i \phi}
\, , \\
{\cal L}_{-+}(\phi)
\!\!\!&=&\!\!\!
\frac{2}{y^2 (1 + \epsilon^2)} \left( 1-y + \frac{\epsilon ^2}{4} y^2 \right) e^{-i 2 \phi}
\, ,
\end{eqnarray}
with the remaining ones related to already found by parity and time-reversal invariance
\begin{eqnarray}
\begin{array}{ll}
{\cal L}_{0 -}(\lambda,\phi)
=
{\cal L}_{0+}(-\lambda,-\phi) \, ,
&
{\cal L}_{\pm,0}(\lambda,\phi)
=
{\cal L}_{0,\pm}(-\lambda,\phi)
\, , \qquad
\\[1mm]
{\cal L}_{--}(\lambda)
=
{\cal L}_{++}(-\lambda) \, ,
&
{\cal L}_{-+} (\phi)
=
{\cal L}_{+-}(-\phi)
\, .
\end{array}
\end{eqnarray}
Using the above relations among helicity amplitudes, we can cast the squared DVCS amplitude
in the following form
\begin{eqnarray}
{\cal Q}^2|{\cal T}^{\rm DVCS}|^2
\!\!\!&=&\!\!\!
2 {\cal L}_{++}(\lambda=0)
\left[T_{++}  \left(T_{+ +}\right)^\ast +T_{+-}  \left(T_{+ -}\right)^\ast \right]
+ 2 {\cal L}_{00}\, T_{0+}  \left(T_{0 +}\right)^\ast
\nonumber\\
&+&\!\!\!
\left[{\cal L}_{0+}(\lambda,\phi)  + {\cal L}_{0+}(-\lambda,-\phi)  \right] T_{0+}
\left[T_{+ +} + T_{+-} \right]^\ast
\nonumber\\
&+&\!\!\!
\left[{\cal L}_{0+}(-\lambda,\phi)  + {\cal L}_{0+}(\lambda,-\phi)  \right]
\left[T_{+ +} +T_{+-} \right]\left(T_{0 +}\right)^\ast
\nonumber\\
&+&\!\!\!
\left[{\cal L}_{+-}(\lambda,\phi) + {\cal L}_{+-}(-\lambda,-\phi)\right]
\left[T_{++}  \left(T_{+-}\right)^\ast +T_{+-}  \left(T_{++}\right)^\ast  \right]
\, .
\end{eqnarray}
These findings immediately allow one to get the Fourier coefficients in the refined
approximation. Substituting the hadronic helicity amplitudes computed to the twist-three
accuracy from Lorentz-covariant parameterizations of the Compton tensor from Section
\ref{LorentzDecomposition},
\begin{eqnarray}
\label{cl-Tpplo}
{\cal T}^{\rm DVCS}_{++}
\!\!\!&=&\!\!\!
{\cal H}+ \mathcal{O} (1/{\cQ}^2)
\, , \\
\label{cal-T0plo}
{\cal T}^{\rm DVCS}_{0+}
\!\!\!&=&\!\!\!
\frac{\sqrt{2}}{2-\Bx} \, \frac{\widetilde{K}}{\cQ} \, {\cal H}_3^{\rm eff}
+ \mathcal{O} (1/{\cQ}^3) +  \mathcal{O} (\alpha_s/{\cQ})
\, , \\
\label{cal-Tmplo}
{\cal T}^{\rm DVCS}_{-+}
\!\!\!&=&\!\!\!
\frac{2}{ (2 - \Bx)^2} \frac{{\widetilde K}^2}{M^2}{\cal H}_{T}
+ \mathcal{O} (1/{\cQ}^2)
\, ,
\end{eqnarray}
where the effective twist-three CFF ${ \cal H}_3^{\rm eff}$ was defined earlier in Eq.~(\ref{Tw3Eff}),
we can read off the kinematically improved DVCS harmonics,
\begin{eqnarray}
\label{Res-Mom-DVCS-0-imp}
c_0^{\rm DVCS}
\!\!\!&=&\!\!\!
2 \frac{2-2 y + y^2+\frac{\epsilon ^2}{2} y^2}{1+\epsilon^2}
{\cal C}^{\rm DVCS} ({\cal H},{\cal H}^\ast;{\cal H}_T,{\cal H}^\ast_T)
+
\frac{16 K^2 {\cal H}_{\rm eff} {\cal H}^\ast_{\rm eff}}{(2-\Bx)^2 (1+\epsilon^2)}
\, , \\
\label{Res-Mom-DVCS-1-imp}
\left\{
{c^{\rm DVCS}_{1} \atop s^{\rm DVCS}_{1}}
\right\}
\!\!\!&=&\!\!\!
\frac{8 K}{{(2 - \Bx)} (1+\epsilon^2)}
\left\{ { (2 - y) \, \Re{\rm e} \atop - \lambda y \sqrt{1 + \epsilon^2} \, \Im{\rm m}} \right\}
\, {\cal C}^{\rm DVCS}
\left(
{\cal H}^{\rm eff};{\cal H}^\ast;{\cal H}_T^\ast\right)
\, , \\
\label{Res-Mom-DVCS-2-imp}
c^{\rm DVCS}_{2}
\!\!\!&=&\!\!\!
\frac{16 {\cal Q}^2 K^2}{M^2 (2 - \Bx)^2 (1+\epsilon^2)}
\Re{\rm e}\,
{\cal C}^{\rm DVCS}_{T} \left( {\cal H}, {\cal H}_T^\ast \right)
\, .
\end{eqnarray}
Here the ${\cal C}$ functions are the same as appeared in Eqs.~(\ref{Def-C-DVCS-unp}). Thus,
to restore the power suppressed contribution in the leptonic part it suffices to perform the
following substitutions, found by comparing Eqs.~\re{Res-Mom-DVCS-0} -- \re{Res-Mom-DVCS-2}
with \re{Res-Mom-DVCS-0-imp} -- \re{Res-Mom-DVCS-2-imp}:
\begin{eqnarray}
\label{Sub-DVCS-0-imp}
2-2 y + y^2
&\Rightarrow&
\frac{2 - 2 y + y^2 + \frac{\epsilon ^2}{2} y^2}{1 + \epsilon^2}\,,
\nonumber\\
\left\{ {2 - y \atop - \lambda y} \right\}
&\Rightarrow&
\frac{1}{1+\epsilon^2} \left\{ {2-y \atop -\lambda y \sqrt{1+\epsilon^2}} \right\}\,,
\\
\frac{{\cal Q}^2 K^2}{M^2}
&\Rightarrow&
\frac{{\cal Q}^2 K^2}{M^2 (1+\epsilon^2)}
\, .
\nonumber
\end{eqnarray}

\subsection{Interference terms}

Finally let us turn to improving the interference $\mathcal{I}$ by treating it in the
manner completely analogous to the previous consideration for the squared DVCS term. As
a result, one finds
\begin{eqnarray}
\label{Def-Sqa-Int-Hel}
{\cal I} =  \frac{\pm e^6}{t \, {\cal P}_1(\phi) {\cal P}_2(\phi)} F (t)
\sum_{a = {\scriptscriptstyle -}, 0, {\scriptscriptstyle +}}
\sum_{b = {\scriptscriptstyle -}, {\scriptscriptstyle +}}
\left\{
{\cal L}^P_{a b}(\lambda,\phi)  T_{a b}
+
\left({\cal L}^P_{a b}(\lambda,\phi) T_{a b}\right)^\ast \right\}
\, ,
\end{eqnarray}
where the leptonic part read
\begin{eqnarray}
{\cal L}^P_{a b}(\lambda,\phi)
=
\epsilon_1^{\mu\ast}(a) L_{\mu\nu\tau}  (p_1 + p_2)^\nu \epsilon_2^\tau(b)
\, ,
\end{eqnarray}
while the definition of the hadronic amplitude is self-obvious. In the present formalism, the
angular dependence is entirely contained in the leptonic part and is rather intricate. Rewriting
the interference in the form
\begin{eqnarray}
{\cal I} &\!\!\!=\!\!\!&  \frac{\pm e^6 F (t)}{t \, {\cal P}_1(\phi) {\cal P}_2(\phi)}
\Bigg[
\left(
{\cal L}^P_{++}+ {\cal L}^P_{--}
\right) T_{+ +}
+
\left(
{\cal L}^P_{0+} + {\cal L}^P_{0-} \right) T_{0 +}
+
\left(
{\cal L}^P_{-+} + {\cal L}^P_{+-}
\right) T_{- +}
+
\mbox{c.c.}
\Bigg] \, ,
\nonumber
\end{eqnarray}
we introduce the decomposition of the leptonic amplitudes in Fourier harmonics encoding the
azimuthal dependence,
\begin{eqnarray}
\mathcal{L}_{{\scriptscriptstyle +} a \, {\scriptscriptstyle +} b}^{P}
+
\mathcal{L}_{{\scriptscriptstyle -} a \, {\scriptscriptstyle -} b}^P
\!\!\!&=&\!\!\!
- \frac{1}{2 \Bx y^3}
\left\{
\sum_{n = 0}^3 C_{ab} (n) \cos (n \phi)
+
i \lambda
\sum_{n = 1}^2 S_{ab} (n) \sin (n \phi)
\right\}
\, .
\end{eqnarray}
The explicit expressions for the exact Fourier coefficients in the leptonic tensor are given in
Appendix \ref{FourierHarmonics}. Notice that once one computes the leptonic part exactly rather
than to the twist-three accuracy, as was done in Section \ref{Sec-BKM}, the simple one-to-one
relation between Fourier coefficients and twist expansion in terms of CFFs is lost beyond the
$1/\cQ^2-$accuracy. This does not prevent one however to project out the real and imaginary
parts of separate CFFs.

\subsection{How robust is leading approximation for harmonics?}
\label{SectionLeptonicTensor}

As we will demonstrate in this section, differences between the approximate and exact amplitudes
can be quite significant due to numerical enhancements of power-suppressed contributions in the
valence and large$-\Bx$ kinematical regions. We ignore the helicity-flip amplitudes in our
consideration and focus our attention on the consequences of the improvements in the twist-two
sector encoded into the approximate Fourier coefficients $c_1^{\mathcal{I}}$ and $s_1^{\mathcal{I}}$
in Eq.\ \re{Res-IntTer-unp}. At first let us propose a ``hot fix'' which is an approximation to
the exact result and accounts for the most significant source of enhanced kinematical corrections,
on the one hand, but leaves dynamical corrections out of the picture, on the other. This constitutes
in the replacements
\begin{eqnarray}
\label{Res-IntTer-unp-c1-hot}
8 K
\left\{  {-(2 - 2y + y^2) \atop \lambda y (2-y)} \right\}
\Rightarrow
\left\{ C_{\pm\pm} (n=1) \atop S_{\pm\pm}(n=1)  \right\}
\end{eqnarray}
in Eq.\ \re{Res-IntTer-unp}. The admixture of higher harmonics proportional to $\cal H$ is not large
for the present experiments, however, one should take care of the zero harmonics by the substitution
in \re{Res-IntTer-unp-c0}
\begin{eqnarray}
\label{Res-IntTer-unp-c0-hot}
- 8  (2 - y)
\Bigg\{
\frac{(2 - y)^2}{1-y} K^2
+
\frac{t}{{\cal Q}^2}  (1-y)(2 - \Bx)
\Bigg\}
\Rightarrow
C_{\pm\pm} (n=0)\,.
\end{eqnarray}
Notice that in this approximation the constant contribution, suppressed by the power of
$1/{\cal Q}$ compared to the first harmonics, is entirely determined by the twist-two CFFs
$\cal H$.

\begin{figure}[t]
\vspace{-2.5cm}
\begin{center}
\mbox{
\begin{picture}(0,293)(250,0)
\put(110,200){(a)}
\put(295,200){(b)}
\put(110,30){(c)}
\put(295,30){(d)}
\put(70,0){\insertfig{13}{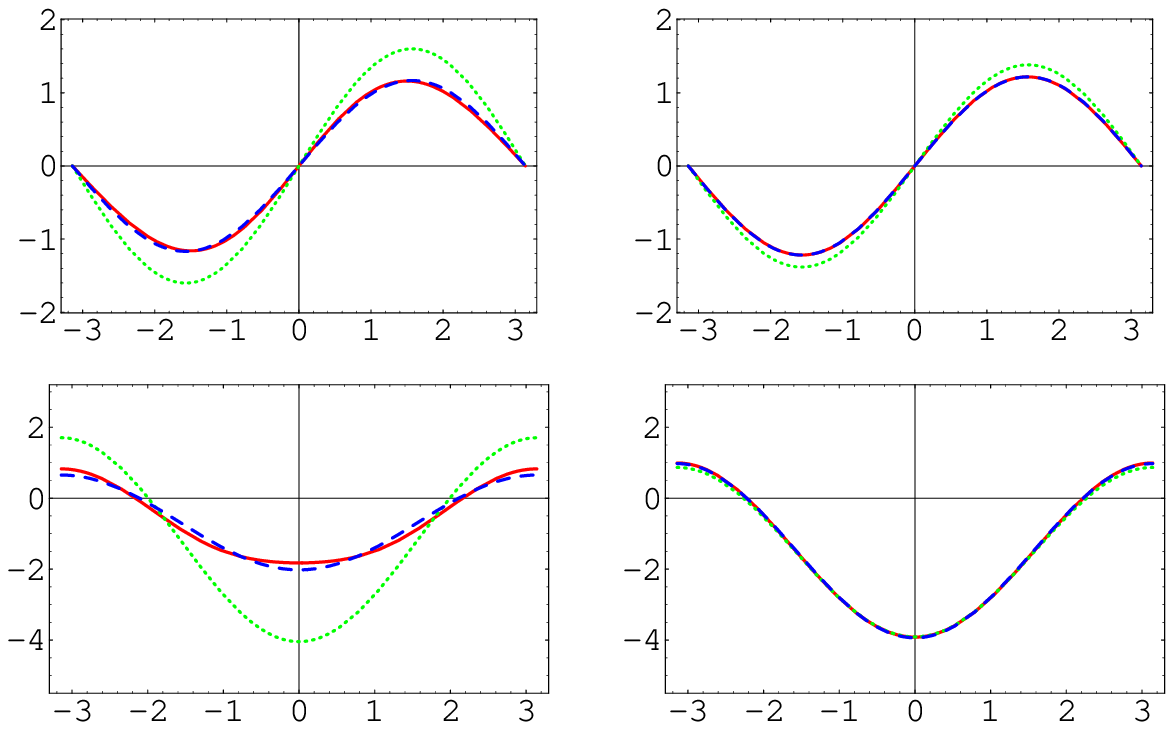}}
\end{picture}
}
\end{center}
\caption{\label{Fig-hotfix} Odd (a,b) and even (c,d) harmonics of the interference term
(\ref{InterferenceTerm}) versus $\phi$ as they arise from the helicity non-flip distribution
amplitude ${\cal T}_{++}$ (set to one) for JLab (a,c) and HERMES (c,d) kinematics as specified
in Eqs.~\re{Kin-JLAB} and \re{Kin-HERMES}, respectively. The dotted and dashed curves emerge
from $s_1 \sin(\phi)$ and $c_0+c_1 \cos(\phi)$ in the approximations of Eqs.\ \re{Res-IntTer-unp-c0}
-- \re{Res-IntTer-unp} and \re{Res-IntTer-unp-c1-hot} -- \re{Res-IntTer-unp-c0-hot}, respectively,
while the solid ones contain all harmonics \re{Res-IntTer-unp-c-im} -- \re{Res-IntTer-unp-s-im}.
}
\end{figure}

This ``hot fix'' provides a significant improvement of the leading approximation, as demonstrated
in Fig.~\ref{Fig-hotfix} where we compared it with exact formulas using for illustrative purposes
the target with the mass $M = 0.94$ GeV . There, in the left and right panels, we present the plots
for the typical kinematical setups, with relatively large values of $t^\prime$, of the present JLab
($E_e= 5.7$ GeV),
\begin{eqnarray}
\label{Kin-JLAB}
t^\prime = -0.3 \, \GeV^2 \, , \quad \Bx = 0.3 \, , \quad \cQ^2 = 1.5 \, \GeV^2
\, ,
\end{eqnarray}
and HERMES ($E_e= 27.5$ GeV),
\begin{eqnarray}
\label{Kin-HERMES}
t^\prime = -0.3 \, \GeV^2 \, , \quad \Bx = 0.1 \, , \ \quad \cQ^2= 2.5\,\GeV^2\,.
\end{eqnarray}
experiments, respectively. A naked eye inspection of the plots of the low$-\mathcal{Q}$/high$-\Bx$
JLab kinematics exhibits the evident feature that the approximation \re{Res-IntTer-unp-c0} and
\re{Res-IntTer-unp} (shown by dotted curves) provides a sizeable false enhancement effect of
the leading twist harmonics compared with the exact result (solid lines). Fortunately, the
contamination by higher harmonics is small and so the substitutions \re{Res-IntTer-unp-c0-hot}
-- \re{Res-IntTer-unp-c1-hot} provide already a good agreement with exact results (dashed line
in Fig.~\ref{Fig-hotfix}). For low$-\cQ^2$/large$-\Bx$, we find $\sim 30\%$ and $\sim 70\%$
deviations for the first odd and even harmonics in the angle $\phi$, respectively. For HERMES,
where $\Bx$ is smaller and $\cQ^2$ larger, the approximations \re{Res-IntTer-unp-c0} and
\re{Res-IntTer-unp} are justified as it is obvious from Fig.~\ref{Fig-hotfix}. We note, however,
that also for this kinematics at larger values of $\Bx$ one may find relative deviations from the
exact results of the order of $15\%$ or so.

One should also be concerned about the size of twist-three effects, originating from the
longitudinal-transverse helicity amplitude $T_{0\pm}$. They are formally suppressed by
$\sqrt{-t^\prime/Q}$ and show up in higher harmonics in the leading approximation of Section
\ref{Sec-BKM}, which, however, are kinematically contaminated by lower ones when computed
exactly. For instance, for the JLab kinematics (\ref{Kin-JLAB}) we find
\begin{eqnarray}
\label{NumericalInter}
{\cal I}
\propto
\!\!\!&-&\!\!\!
1.34 \left[ \cos(\phi) + 0.51 - 0.14 \cos(2\phi) \right] {\Re}{\rm e} \, {\cal H}
-
0.68 \left[ \cos(2\phi) - 0.32 + 0.21 \cos(\phi) \right] {\Re}{\rm e} \, {\cal H}_3^{\rm eff}
\nonumber \\
&+&\!\!\!
1.16 \lambda
\left[ \sin(\phi) + 0.04 \sin(2\phi) \right] {\Im}{\rm m} \, {\cal H}
+
0.23 \lambda \left[ \sin(2\phi) + 0.79 \sin(\phi) \right] {\Im}{\rm m} \, {\cal H}_3^{\rm eff}
\, .
\end{eqnarray}
If the twist-tree CFF ${\cal H}_3^{\rm eff}$ is comparable in magnitude to the twist-two
${\cal H}$, it may provide a sizeable contribution to the lower harmonics, i.e., $\sin(\phi)$,
$\cos(\phi)$ and the constant part. The size of the $\sin(2\phi)$ and $\cos(2\phi)$ harmonics
will serve as an accurate estimate of this admixture. Note that for (very) small values of
$\sqrt{-t^\prime/Q}$, the twist-two and -three harmonics start to decouple. However, even for
HERMES kinematics one must account for these mixing effects for average values of
$-t \sim 0.3 \, \GeV^2$.

Finally, we mention that in the present estimate \re{NumericalInter} we ignored the
transverse-transverse helicity-flip amplitude $T_{+-}$ since it safely decouples from the other
ones. It shows up in the $\cos(3\phi)$ harmonics, which is only slightly affected by the
twist-two and -three CFFs. On the other hand, its contribution to the lower harmonics is negligible.

Since the ultimate goal of the exclusive cross section \re{WQ} measurements is the extraction of CFFs
and consequently clean separation of GPDs, which shed light on the internal structure of hadrons, data
analyses have to disentangle the intricate dependence of kinematical factors on dynamical variables
from the one of GPDs themselves. Therefore, the role of $1/\cal Q$ power suppressed contributions
should not be underestimated as these enter dressed with numerically enhanced factors being function
of $y$, $\Bx$ and $t$. Therefore, for the most robust CFF extraction, we suggest to utilize the
following general formulas for the Fourier coefficients in data analyses
\begin{eqnarray}
\label{Res-IntTer-unp-c-im}
c^{\cal I}_{n}
\!\!\!&=&\!\!\!
C_{++} (n) \, \Re{\rm e} \, {\cal C}^{\cal I} \left({\cal H}\right)
+
\frac{\sqrt{2}}{2-\Bx}\, \frac{\widetilde{K}}{\cQ}
\, C_{0+} (n) \, \Re{\rm e} \, {\cal C}^{\cal I} \left( {\cal H}^{\rm eff}_3 \right)
+
\frac{2}{(2-\Bx)^2} \, \frac{\widetilde{K}^2}{M^2}
\, C_{-+} (n) \, \Re{\rm e} \, {\cal C}^{\cal I} \left( {\cal H}_T \right)
\, , \nonumber\\[-3mm]
&& \\
\label{Res-IntTer-unp-s-im}
s^{\cal I}_{n}
\!\!\!&=&\!\!\!
S_{++} (n) \, \Im{\rm m} \, {\cal C}^{\cal I} \left({\cal H}\right)
+
\frac{\sqrt{2}}{2-\Bx}\, \frac{\widetilde{K}}{\cQ}
\, S_{0+} (n) \, \Im{\rm m} \, {\cal C}^{\cal I} \left({\cal H}^{\rm eff}_3\right)
+
\frac{2}{(2-\Bx)^2} \, \frac{\widetilde{K}^2}{M^2}
\, S_{-+} (n) \, \Im{\rm m} \, {\cal C}^{\cal I} \left({\cal H}_T\right)
\, , \nonumber\\[-3mm]
&&
\end{eqnarray}
in the interference term \re{InterferenceTerm}. Though this results into a more involved
extraction procedure, compared to that one described in Ref.~\cite{BelMueKir01}, it will
not introduce any complications of principle.

\section{Uncertainties arising from the twist expansion}
\label{SectionHadronicTensor}

There are various possibilities to write down the tensor decomposition for the DVCS tensor
\re{DVCSamplitude} respecting its gauge invariance and Lorentz covariance. Each of them
will be associated with a specific set of CFFs. Of course, all parameterizations are
equivalent. However, relating physical CFFs to partonic GPDs requires an expansion in
inverse powers of the hard scale $\mathcal{Q}$, known as the twist expansion. As a
consequence, expressions obtained for different parameterizations will only be equivalent
to the considered order in the $1/\mathcal{Q}-$expansion and differ when higher order
terms are taken into account. Let us address several Lorentz structures used in the
literature and discuss their manifestation in physical observables.

Presently, the hadronic Compton amplitude \re{DVCSamplitude} has been worked
out to the twist-three accuracy within the GPD formalism
\cite{AniPirTer00,PenPolShuStr00,BelMue00a,RadWei00,KivPolSchTer00}. While the leading
twist sector is known in the next-to-leading \cite{JiOsb97} and next-to-next-to-leading
\cite{Mul05} orders in the modified minimal subtraction and the special conformal
subtraction schemes, respectively, the twist-three coefficient functions are available
in the handbag approximation only (with a partial one-loop calculation reported in
Ref.~\cite{KivMan03}). The restriction of the current analyses to account for $1/\mathcal{Q}$
suppressed effects is a drawback which yields insufficient accuracy to render the DVCS tensor
gauge invariant and obey the transversality condition, arising from the current conservation,
\begin{eqnarray}
\label{Con-Tra}
q_{1}^\nu T_{\mu\nu} = q_{2}^\mu T_{\mu\nu} \equiv 0 \, ,
\end{eqnarray}
exactly. To illustrate the interplay of higher twist contributions and the restoration of
the transversality condition of the DVCS tensor, let us recall the role of the twist-three
effects in this endeavor. For a scalar target, the twist-two result in the handbag approximation
reads:
\begin{eqnarray}
\label{Def-Ttree}
T^{(2)}_{\mu\nu}
=
- \frac{1}{p \cdot q}
\left[ (p \cdot q) g_{\mu\nu} - q_{\mu} p_{\nu} - q_{\nu} p_{\mu} - \xi p_{\mu} p_{\nu} \right]
{\cal H} (\xi, t; {\cal Q}^2)
\, ,
\end{eqnarray}
where we used the symmetric variables $p = p_1 + p_2$ and $q = \ft12 (q_1 + q_2)$ which spawn
the generalized Bjorken variable $\xi = Q^2/p \cdot q$ defined in terms of the average photon
virtuality $Q^2 = -q^2$. In the non-forward kinematics of DVCS, obviously this tensor does not
respect the current conservation and so one finds after contraction, e.g., with $q_{2}^\mu$,
that the transversality condition \re{Con-Tra} is violated by $\Delta^\perp_{\nu} = \Delta_{\nu}
- \eta p_{\nu}$ which, according to the power counting, is a twist-three effect. The variable
$\eta$ is the $t$-channel scaling variable known as skewness $\eta = (\Delta \cdot q)/(p \cdot q)$.
It is related to the generalized Bjorken variable $\xi$ for the DVCS kinematics via the relation
$\eta = -\xi/(1+ t/2{\cQ^2}) \approx - \xi$. Evaluation of the twist-three amplitudes yields
beside a contribution proportional to the function $\mathcal{H}_3$
\begin{eqnarray}
\label{Def-Ttree-3}
T^{(3)}_{\mu\nu} =
\frac{1}{p \cdot q}
\left[
\Delta_\mu^\perp \left( q_{2\nu}+ \xi p_\nu \right)
+
\Delta_\nu^\perp \left( q_{1\mu}+ \xi p_\mu \right)
\right]
{\cal H}_3
\, ,
\end{eqnarray}
also an antisymmetric tensor accompanied by the twist-two CFF $\mathcal{H}$
$$
\delta T^{(2)}
=
\frac{1}{p \cdot q}
\left[
\Delta^\perp_{\mu} p_{\nu} - \Delta^\perp_{\nu} p_{\mu}
\right]
{\cal H}
\, .
$$
The latter when combined with Eq.~(\ref{Def-Ttree}), $T^{(2)} + \delta T^{(2)}$, generates the
replacement of the twist-two Lorentz structure with
$$
- (p \cdot q) g_{\mu\nu} + q_{\mu} p_{\nu} + q_{\nu} p_{\mu} + \xi p_{\mu} p_{\nu}
\rightarrow
- (p \cdot q) g_{\mu\nu} + q_{1\mu} p_{\nu} + q_{2\nu} p_{\mu} + \xi p_{\mu} p_{\nu}
\, .
$$
It is straightforward to check now that the violation of the transversality condition is
postponed to the twist-four level, i.e., $\mathcal{O} (\mathcal{Q}^{-2})$. Thus addressing
it requires a full-fledged twist-four analysis along the lines of Ref.\ \cite{EllFurPet83}.
Since such a calculation has not been performed yet in a consistent manner, we will take
a pragmatic point of view and restore the exact gauge invariance in several superficially
inequivalent ways and study the resulting numerical differences.

\subsection{Lorentz decomposition of the Compton tensor}
\label{LorentzDecomposition}

The  complete twist-three result for the DVCS tensor for a scalar
target has three independent CFFs, which is the maximal possible
number allowed by the underlying symmetries. The result reads in
terms of physical four-momenta as follows
\begin{eqnarray}
\label{Ten-cal}
T^{(2+3)}_{\mu \nu} \!\!\!&=&\!\!\!
-
\frac{1}{(p \cdot q)}
\left[
(p \cdot q) g_{\mu\nu} - q_{1\mu} p_{\nu} - q_{2\nu} p_{\mu} - \xi p_{\mu} p_{\nu}
\right] \left( {\cal H} + \frac{\Delta_\perp^2}{2 M^2} {\cal H}_{\rm T} \right)
\nonumber\\
&&\!\!\!
+ \,
\frac{1}{M^2} \Delta_{\perp \mu} \Delta_{\perp \nu}
{\cal H}_{\rm T}
+ \frac{1}{(p \cdot q)}
\left[
\Delta_{\perp \mu} \left(q_{2\nu}+ \xi p_\nu \right)
+
\Delta_{\perp \nu} \left(q_{1\mu}+ \xi p_\mu\right)
\right] {\cal H}_{3}
\, .
\nonumber
\end{eqnarray}
As mentioned above the two  Lorentz tensors, proportional to the CFFs $\cal H$ and
${\cal H}_{3}$, are transverse up to twist-four terms.  The third structure is a
symmetric and traceless tensor, which is induced by the gluon transversity (photon
helicity flip by two units) \cite{HooJi98,BelMue00,Die01}. It is only transversal
to the order $\Delta_\perp/M$.

Utilizing the equation of motion at twist-four level, should allow one to satisfy the
transversality condition (\ref{Con-Tra}) exactly, since higher than geometric twist-four
contribution are formally absent in the handbag diagram being proportional to derivatives
of Dirac delta function. The restoration of the transversality condition for the Lorentz
structures, proportional to ${\cal H }$ and ${\cal H}_{3}$ requires a $t-$dependent term,
since neglecting such terms results in a failure of (\ref{Con-Tra}) \cite{BelMue01}. The
``minimalist'' restoration in the twist-two sector requires to add a $(t/Q^2) p_\mu p_\nu$
contribution, leading to the replacement
\begin{eqnarray}
\label{Rec-GauInv1}
(p \cdot q) g_{\mu\nu} - q_{1\mu} p_{\nu} - q_{2\nu} p_{\mu} - \xi p_{\mu} p_{\nu}
\quad \rightarrow \quad
(p \cdot q) g_{\mu\nu} - q_{1\mu} p_{\nu} - q_{2\nu} p_{\mu} -
\left( 1 + \frac{t}{4 Q^2} \right) \xi p_{\mu} p_{\nu}
\, .
\end{eqnarray}
In the twist-three sector one needs a
\baa
\frac{t}{Q^2}\frac{q_{\mu} p_{\nu} - q_{\nu} p_{\mu} + \eta p_\mu p_\nu}{p \cdot q}
\eaa
proportional term, which can be absorbed by adding power suppressed contributions to the
following four-vectors:
\begin{eqnarray}
\label{Rec-GauInv2}
\Delta_{\perp\mu} \rightarrow \Delta_{\perp\mu} \pm \frac{t}{2 Q^2} \xi p_\mu
\, , \qquad
q_{i \mu} + \xi p_\mu \rightarrow  q_{i \mu}+ \left( 1 + \frac{t}{4 Q^2} \right) \xi p_\mu
\, ,
\end{eqnarray}
with $\pm$ sign in the first equation standing for indices associated with outgoing/incoming
photons. The restoration of the gauge invariance for the gluon transversity is more cumbersome,
since it will already affected at the twist-three level. It can be achieved by adding a
$\Delta_\perp$ proportional terms \cite{KivMan01}. For later illustration we adopt here the
recipes (\ref{Rec-GauInv1}) and (\ref{Rec-GauInv2}). Finally, the DVCS tensor can then be
written as
\begin{eqnarray}
\label{Ten-imp}
T_{\mu \nu}
=
\!\!\!&-&\!\!\!
\frac{1}{p \cdot q}
\left[
(p \cdot q) g_{\mu\nu} - q_{1\mu} p_{\nu} - q_{2\nu} p_{\mu}
-
\left(1 + \frac{t}{4 Q^2} \right) \xi p_{\mu} p_{\nu}
\right] \left( {\cal H} + \frac{\Delta_\perp^2}{2 M^2} {\cal H}_{\rm T} \right)
\\
&+&\!\!\!
\frac{1}{M^2}
\left(\Delta_{\perp\mu} + \frac{t}{2 Q^2} \xi p_\mu\right)
\left(\Delta_{\perp\nu} - \frac{t}{2 Q^2} \xi p_\nu\right)
{\cal H}_{\rm T}
\nonumber\\
&+&\!\!\!
\frac{1}{p \cdot q}
\Bigg[
\left(
q_{2\nu}+ \left(\!1+ \frac{t}{4 Q^2} \!\right) \xi p_\nu
\right)\!
\left(
\Delta_{\perp\mu} + \frac{t}{2 Q^2} \xi p_\mu
\right)
\nonumber\\
&&\qquad\qquad\qquad\qquad\qquad\quad\
+
\left(
q_{1\mu}+ \left(\!1+ \frac{t}{4Q ^2} \!\right)  \xi p_\mu
\right)\!
\left(
\Delta_{\perp\nu} -\frac{t}{2 Q^2} \xi p_\nu
\right)
\Bigg]
{\cal H}_{3}
\, .
\nonumber
\end{eqnarray}
For DVCS kinematics this parameterization is complete and, thus, it can  be uniquely mapped into a
different form, e.g., used in Ref.~\cite{BelKirMueSch01}.

Another way to restore gauge invariance of the amplitude \re{DVCSamplitude} beyond twist-three
accuracy is by introducing a projector \cite{BelMue00a}
\be
{\cal P}^{\mu\nu}
\equiv
g^{\mu\nu}
-
\frac{q_1^\mu \ q_2^\nu}{q_1 \cdot q_2}
\, ,
\ee
fulfilling the conditions ${\cal P}^{\mu\nu} q_{1\nu} = q_{2\mu} {\cal P}^{\mu\nu} = 0$. They
provide a transverse Compton tensor when contracted on both sides with the twist-two DVCS amplitude
$T^{(2)}_{\mu\nu}$. When expanded to twist-three accuracy this procedure reproduces gauge-restoring
terms coinciding with the ones obtained from the explicit analysis reviewed in the preamble to
Section \ref{SectionHadronicTensor}. Notice however that contrary to the consideration in the
previous paragraph, this recipe generates an infinite tower of higher-twist contributions when
expanded in inverse powers of the average photon virtuality.

Finally, let us address the formalism of Ref.\ \cite{VanGuiGui99}. In applications of the QCD-improved
parton model to high-energy scattering, it is customary to parameterize the Compton tensor in terms of
the light-cone vector $n^\mu$ and its tangent $\widetilde n^\mu$ such that $n^2 = \widetilde n^2 = 0$
and $n \cdot \widetilde n = 1$. In particular, these can be chosen as the plus and minus components of
the initial photon and proton momenta, respectively, cf.\ Fig.~\ref{Fig-Kin}. However, the choice of the
light-cone vectors is not unique, since it implicitly refers to a given reference frame. For instance,
the parameterization used in Ref.\ \cite{VanGuiGui99},
\begin{eqnarray}
\label{Var-Phy2LC}
q_{1\mu}
=
\frac{\cQ^2}{4\xi^\prime} n_\mu - 2 \xi^\prime \widetilde n_\mu
\, , \quad
p_\mu
=
2 \widetilde n_\mu + \frac{4 M^2 - t}{2} n_\mu
\, , \quad
\Delta_\mu
=
- (2\bar{\xi}) \widetilde n_{\mu} + \bar{\xi}\, \frac{4 M^2 - t}{4} n_\mu + \bar{\Delta}_{\perp\mu}
\, ,
\end{eqnarray}
is done in a ``collinear'' frame ($\vec{p}_{1 \perp} = - \vec{p}_{2 \perp}$). We note that the
scaling variables $\xi^\prime$ and $\bar{\xi}$ are proportional to the generalized Bjorken variable
$\xi$ and skewness $\eta \simeq - \xi$, respectively, but, differ from them by power suppressed
corrections. The DVCS tensor parameterized in terms of the light-cone vectors rather than physical
four-vectors reads \cite{VanGuiGui99}
\begin{eqnarray}
\label{ten-VGG}
T_{\mu\nu} =
\!\!\!&-&\!\!\!
\left[g_{\mu\nu}- n_\mu \widetilde n_\nu - \widetilde n_\mu n_\nu +
\frac{\widetilde n_\mu  \bar{\Delta}_{\perp\nu}}{\widetilde n \cdot q_2} \right] {\cal H}
\nonumber\\
&+&\!\!\!
\left[
\bar{\Delta}_{\perp\mu} + \bar{\Delta}_\perp^2 \frac{\widetilde n_\mu }{\widetilde n \cdot q_2}
\right]
\left[
\frac{n_{\nu}}{2} + 4 \xi^{\prime 2} \frac{\widetilde n_\nu}{{\cal Q}^2}
\right] {\cal H}_{3}
\\
&+&\!\!\!
\left[
\frac{\bar{\Delta}_{\perp\mu} \bar{\Delta}_{\perp\nu} }{M^2}
-
\frac{\bar{\Delta}_\perp^2}{2 M^2}
\left(g_{\mu\nu}- n_\mu \widetilde n_\nu - \widetilde n_\mu n_\nu -
\frac{\widetilde n_\mu  \bar{\Delta}_{\perp\nu} }{\widetilde n \cdot q_2} \right)
\right] {\cal H}_{\rm T}
\, .
\nonumber
\end{eqnarray}
In the first and third Lorentz structure the linear term $\bar{\Delta}_{\perp\nu}$ exactly restores
the transversality, while for the second Lorentz structure  a $\bar{\Delta}^2-$proportional term is
needed. Note that the restoration of transversality for the gluon transversity-induced Lorentz
structure essentially differs from the prescription employed in the parameterization (\ref{Ten-imp}).
In the frame we are adopting, the transversal helicity amplitudes, $\gamma^\ast_T \to \gamma_T$, are
given by CFFs ${\cal H}$ and ${\cal H}_T$, while the the longitudinal one, $\gamma^\ast_L\to \gamma_T$,
is entirely related to the twist-three CFF ${\cal H}_{3}$.

To get rid of the frame dependence in the parameterization (\ref{ten-VGG}), we use the relations
(\ref{Var-Phy2LC}) to express the DVCS tensor in terms of the physical momenta. After this, the
tensor takes a frame-independent Lorentz covariant form. Comparing it to the parameterization
(\ref{Ten-imp}) we can read off a rather cumbersome relation among the two sets of CFFs.

\subsection{Numerical estimates}

The leading contribution to each hadronic helicity amplitude from any tensor decomposition,
either \re{Ten-cal} or \re{Ten-imp} or \re{ten-VGG}, is universal and is given by Eqs.\
\re{cl-Tpplo} -- \re{cal-Tmplo}. Differences will arise starting with $1/\mathcal{Q}^2$
contributions. To illustrate uncertainties related to the twist-four effects we evaluate
the helicity amplitudes to the order $1/{\cQ}^3$. The differences in the helicity conserved
amplitude from the calculated  (\ref{Ten-cal}) and improved (\ref{Ten-imp}) DVCS tensor is
\begin{eqnarray}
{\cal T}^{\rm DVCS}_{++}\big|_{{\rm Eq.(\ref{Ten-cal})}}
-
{\cal T}^{\rm DVCS}_{++}\big|_{{\rm Eq.(\ref{Ten-imp})}}
\!\!\!&\simeq&\!\!\!
\mathcal{O} \left( \frac{\Bx^2 t\, t^\prime}{\cQ^4} \right)
\, .
\end{eqnarray}
It can be considered as negligible and plays practically no role. The ambiguities in the
restoration of transversality with \re{Ten-cal} shows up mainly in the gluon transversity
induced sector yielding
\be
{\cal T}^{\rm DVCS}_{++}\big|_{{\rm Eq.(\ref{ten-VGG})}}
-
{\cal T}^{\rm DVCS}_{++}\big|_{{\rm Eq.(\ref{Ten-imp})}}
=
- \frac{(4 M^2 - t) \Bx^3}{\cQ^2} \frac{4 (1 - \Bx) \Bx M^2 - (4 - 3\Bx) t}{(2 - \Bx)^4 M^2 } {\cal H}_T
+
\mathcal{O} (1/\cQ^4)
\, .
\ee
It is suppressed for small $\Bx$ by a factor $\Bx^3$. The CFFs $\cal H$ and ${\cal H}_3$ enter
here as a $\Bx^2 t t^\prime/\cQ^4$ and $\Bx^4 M^2 t^\prime/\cQ^4$ suppressed contributions, which
are practically very small.

For the longitudinal-transverse helicity amplitude, the higher twist effects are more pronounced
and yield the difference
\begin{eqnarray}
{\cal T}^{\rm DVCS}_{0+}\big|_{{\rm Eq.(\ref{Ten-cal})}}
\!\!\!&-&\!\!\!
{\cal T}^{\rm DVCS}_{0+}\big|_{{\rm Eq.(\ref{Ten-imp})}}
\\
&=&\!\!\!
\frac{ \Bx \widetilde{K} t}{2 \sqrt{2} (2 - \Bx) \cQ^3}
\left(
{\cal H} \left[1 + \mathcal{O} (1/\cQ^2) \right]
-
2 \frac{1 - 2 \Bx}{2 - \Bx} {\cal H}_3 \left[ 1 + \mathcal{O} (1/\cQ^2) \right]
\right)
\nonumber\\
&+&\!\!\!
\frac{
\widetilde{K}}{\cQ} \frac{2 M^2 \Bx^2 - t \left(2 - 2 \Bx + \Bx^2\right)
}{
\sqrt{2} M^2 (2 - \Bx)^3} {\cal H}_{\rm T} \left[1 + \mathcal{O} (1/\cQ^2) \right]
\, . \nonumber
\end{eqnarray}
While the two recipes (\ref{Ten-imp}) and (\ref{ten-VGG}) yields $\Bx (4 M^2 - t){\cal H}_3^{\rm eff}
/4\cQ^2$, $\Bx^2(4M^2-t){\cal H}^{\rm eff} /4\cQ^2$ and $\Bx^2(4M^2-t){\cal H}_{\rm T}/4\cQ$ suppressed
differences.

Finally, for the transverse-transverse helicity-flip amplitude, we find that the ambiguities due to
kinematical higher-twist corrections enter at order $1/\cQ^4$ for the recipe (\ref{Ten-imp}), i.e.,
\begin{eqnarray}
{\cal T}^{\rm DVCS}_{-+}\big|_{{\rm Eq.(\ref{Ten-cal})}}
-
{\cal T}^{\rm DVCS}_{-+}\big|_{{\rm Eq.(\ref{Ten-imp})}}
\sim O(1/\cQ^4)
\, ,
\end{eqnarray}
however, only at order $1/\cQ^2$ for the prescription (\ref{ten-VGG}).

To give a numerical example demonstrating the contamination of the leading contribution by the
ambiguities of the power suppressed effects, let us start with the JLab kinematics (\ref{Kin-JLAB}).
Numerically, we find that the non-flip helicity amplitude can be safely approximated by the CFF
$\cal H$, cf.~Eq.~(\ref{cl-Tpplo}):
\begin{eqnarray}
{\cal T}^{\rm DVCS}_{++}
=
\left\{\!
\begin{array}{r}
0.997 \\
0.996 \\
1.003
\end{array}
\!\right\}
{\cal H}
+
\left\{\!
\begin{array}{l}
0.010 \\
0.011 \\
0.008
\end{array}
\!\right\}
{\cal H}_3
+
\left\{\!
\begin{array}{l}
0.019 \\
0.019 \\
0.000
\end{array}
\!\right\}
{\cal H}_T
\quad \mbox{for} \quad
\left\{\!
\begin{array}{l}
{\rm Eq.~(\ref{Ten-cal})} \\
{\rm Eq.~(\ref{Ten-imp})} \\
{\rm Eq.~(\ref{ten-VGG})}
\end{array}
\!\right\}
\, .
\end{eqnarray}
Certainly, here we can practically set ${\cal T}^{\rm DVCS}_{++} = {\cal H}$. Unfortunately, for the
longitudinal-transverse helicity flip amplitude we find rather strong deviations from the leading
approximation (\ref{cal-T0plo}) that are caused by kinematical corrections:
\begin{eqnarray}
\frac{ (2-\Bx)  \cQ {\cal T}^{\rm DVCS}_{0+}}{\sqrt{2}\widetilde{K} }
=
\left\{\!
\begin{array}{l}
1.30 \\
1.34 \\
0.91
\end{array}
\!\right\}
{\cal H}_3^{\rm eff}
+
\left\{\!
\begin{array}{r}
-0.21 \\
-0.17 \\
0.02
\end{array}
\!\right\}
{\cal H}
-
\left\{\!
\begin{array}{l}
0.16 \\
0.34 \\
0.03
\end{array}
\!\right\}
{\cal H}_T
\quad \mbox{for} \quad
\left\{\!
\begin{array}{l}
{\rm Eq.~(\ref{Ten-cal})} \\
{\rm Eq.~(\ref{Ten-imp})} \\
{\rm Eq.~(\ref{ten-VGG})}
\end{array}
\!\right\}
\, . \nonumber\\
\end{eqnarray}
Although the effect of restoration of the transversality from the twist-two amplitude \re{Ten-cal}
is small within the recipe (\ref{Ten-imp}) (except for the gluon transversity), the numerical
values deviate considerably for (\ref{ten-VGG}). We suggest to rely for simplicity on the leading
approximation (\ref{cal-T0plo}). Finally, for the transverse-transverse
helicity flip amplitude we again observe that the deviations from the leading approximation
(\ref{cal-Tmplo}) are negligible except for the parameterization (\ref{ten-VGG}):
\begin{eqnarray}
\frac{ (2-\Bx)^2  M {\cal T}^{\rm DVCS}_{-+}}{2\widetilde{K}^2 }
=
\left\{\!
\begin{array}{r}
1.01 \\
1.00 \\
0.81
\end{array}
\!\right\}
{\cal H}_{\rm T}
+
\left\{\!
\begin{array}{r}
-0.02 \\
-0.02 \\
0.02
\end{array}
\!\right\}
{\cal H}
-
\left\{\!
\begin{array}{l}
0.06 \\
0.07 \\
0.04
\end{array}
\!\right\}
{\cal H}_3
\quad \mbox{for} \quad
\left\{\!
\begin{array}{l}
{\rm Eq.~(\ref{Ten-cal})} \\
{\rm Eq.~(\ref{Ten-imp})} \\
{\rm Eq.~(\ref{ten-VGG})}
\end{array}
\!\right\}
\, . \nonumber\\
\end{eqnarray}

It is rather obvious that for decreasing $\Bx$ and/or increasing $\cQ^2$ the ``kinematical" power
corrections are getting reduced. For instance, for HERMES kinematics (\ref{Kin-HERMES}) we find
for the most problematic longitudinal-transverse helicity-flip amplitude
\begin{eqnarray}
\frac{ (2-\Bx)  \cQ {\cal T}^{\rm DVCS}_{0+}}{\sqrt{2}\widetilde{K} }
=
\left\{\!
\begin{array}{l}
1.00 \\
1.03 \\
1.01
\end{array}
\!\right\}
{\cal H}_3^{\rm eff}
+
\left\{\!
\begin{array}{r}
-0.02 \\
-0.01 \\
0.00
\end{array}
\!\right\}
{\cal H}
+
\left\{\!
\begin{array}{r}
0.05 \\
-0.04 \\
-0.02
\end{array}
\!\right\}
{\cal H}_T
\quad \mbox{for} \quad
\left\{\!
\begin{array}{l}
{\rm Eq.~(\ref{Ten-cal})} \\
{\rm Eq.~(\ref{Ten-imp})} \\
{\rm Eq.~(\ref{ten-VGG})}
\end{array}
\!\right\}
\, . \nonumber\\
\end{eqnarray}
This exhibits the legitimacy of the leading approximation \re{cal-T0plo} employed in Refs.\
\cite{BelMueKir01,BelMueKirSch00}.

\section{Conclusions}

The main goal of the present consideration was understanding of the power-suppressed effects
in DVCS observables stemming from the exact account for kinematical contributions in the
hadronic mass $M^2$ and momentum transfer $t$. Using the photon helicity amplitudes, we
separated the leptoproduction cross section in terms of the leptonic and hadronic helicity
amplitudes. The choice of the target rest frame with the $z-$axis directed (counter)along
to the virtual photon three-momentum allowed one to localize its dependence on the azimuthal
angle to the leptonic part. These were then computed exactly to leading order in QED fine
structure constant thus improving approximate results of previous considerations
\cite{BelMueKir01,BelMueKirSch00}.

Numerical estimates performed for the current kinematics of JLab experiments demonstrated
that, due to rather low virtuality of the hard photon and valence-region values for the
Bjorken variable, the restriction to merely the leading approximation of Ref.\
\cite{BelMueKir01,BelMueKirSch00} yields significant overestimate of event rates compared
to the exact treatment. However, for higher values of the hard scale, typical for the HERMES
experiment, the approximation of the earlier work becomes legitimate. We proposed a set of
formulas for refined analysis of DVCS observables. Although, in the improved approximation
the classification scheme of Ref.\ \cite{BelMueKir01}, according to which the Fourier
harmonics are strictly associated with the twist of the contributing CFFs, is altered, this
does not represent a difficulty of principle to extract CFFs from experimental observables.
However, obviously the inversion problem becomes more tedious.

Let us point out that the choice of Lorentz invariant kinematical variables in the evaluation
of CFFs from the corresponding GPDs is also not unique as it has a cross talk with higher-twist
contributions. The optimal choice should minimalize $1/\cQ^2-$suppressed contributions. This
problem was not the focus of our present analysis, where we used as in our previous studies a
legitimate choice $\xi = - \eta = \Bx/(2 - \Bx)$ and took the photon virtuality $\cQ^2$ as the
large scale.

The consideration of the present work can be generalized in a straightforward fashion to
targets possessing nonvanishing spin, with nucleon being the most interesting one. We
anticipate that our current analysis will quantitatively hold for DVCS off an unpolarized
proton target too. The ``hot fixes'' \re{Sub-DVCS-0-imp}, \re{Res-IntTer-unp-c1-hot}, and
\re{Res-IntTer-unp-c0-hot} can be immediately used to improve on Eqs.\ (43) -- (45), (54),
and (53), respectively, of Ref.~\cite{BelMueKir01}. Note that for the spin-one-half target
new combinations of CFFs will emerge, e.g., $\Delta C$ in the interference term. However,
we expect that such contributions, induced by the helicity flip of the outgoing proton, are
relatively unimportant for the kinematics of present experiments. We will report on this
analysis in future work.

\vspace{0.5cm}

\noindent We are indebted to H.~Avakian, M.~Gar\c{c}on, M.~Guidal, and F.~Sabati\'{e} for
discussions which initiated our studies. This work was supported by the U.S. National
Science Foundation under grants no.~PHY-0456520 and no.~PHY-0757394 and funds provided by
the ASU College of Liberal Arts and Sciences. D.M.~would like to thank the Particle Physics
and Astrophysics Group at ASU for hospitality extended to him during the final stage of the
work.

\appendix

\section{Kinematics in the target rest frame}
\label{KinematAppend}

We fix our kinematics by going to the target rest frame and choosing the $z$-component of
the virtual photon momentum negative and the $x$-component of the incoming lepton being
positive. The components of the corresponding four-vectors read
\be
p_1
= (M, 0, 0, 0)
\, , \quad
q_1
=
\frac{\mathcal{Q}}{\epsilon}
\left(1, 0, 0, - \sqrt{1 + \epsilon^2}
\right)
\, , \quad
k
=
\frac{\mathcal{Q}}{y \epsilon}
\left(
1 , \sin\theta_l, 0, \cos\theta_l
\right) \, ,
\ee
with the lepton scattering angle being
\be
\cos \theta_l = - \frac{1 + \frac{y \epsilon^2}{2}}{\sqrt{1 + \epsilon^2}}
\, , \qquad
\sin \theta_l = \frac{\epsilon \sqrt{ 1 - y -  \frac{y^2 \epsilon^2}{4}}}{\sqrt{1 + \epsilon^2}}
\, .
\ee
The outgoing momenta are parameterized in terms of the scattering angles in the hadronic plane,
see Fig.\ \ref{Fig-Kin},
\ba
q_2
\!\!\!&=&\!\!\!
\frac{\mathcal{Q}^2 + \Bx t}{2 M \Bx}
\left(
1 ,
\cos\varphi_\gamma \sin\theta_\gamma ,
\sin\varphi_\gamma \sin\theta_\gamma,
\cos\theta_\gamma
\right) \, , \\
p_2
\!\!\!&=&\!\!\!
\left(
M - \frac{t}{2 M}, \sqrt{- t + \frac{t^2}{4 M^2}} \cos\phi \sin\theta ,
\sqrt{- t + \frac{t^2}{4 M^2}} \sin\phi \sin\theta, \sqrt{- t + \frac{t^2}{4 M^2}}
\cos\theta
\right) \, ,
\ea
where the polar angles read in terms of the kinematical variables of the phase space
\be
\cos \theta_\gamma
=
-
\frac{1 + \frac{\epsilon^2}{2} \frac{\mathcal{Q}^2 + t}{\mathcal{Q}^2 + \Bx t}}{\sqrt{1 + \epsilon^2}}
\, , \qquad
\cos \theta
= -
\frac{\epsilon^2 (\mathcal{Q}^2 - t) - 2 \Bx t}{4 \Bx M \sqrt{1 + \epsilon^2} \sqrt{- t + \frac{t^2}{4 M^2}}}
\, .
\ee
The azimuthal angle of the photon $\varphi_\gamma$ is related to the one of the outoing hadron $\phi$
via $\varphi_\gamma = \phi + \pi$.

The explicit component form of the photon four-momenta allows one to construct their polarization vectors:
\begin{eqnarray}
\label{Polarization1}
\varepsilon_1^\mu(\pm)
\!\!\!&=& \!\!\!
\frac{e^{\mp i\phi}}{\sqrt{2}}(0,1, \pm i,0)\,, \qquad
\varepsilon_1^\mu(0) = \frac{{\cal Q}}{\sqrt{2 \Bx M}}(-\sqrt{1+\epsilon^2},0, 0,1)
\, , \\
\label{Polarization2}
\varepsilon_2^{\mu\ast}(\pm)
\!\!\!&=&\!\!\!
\frac{1}{\sqrt{2}}
\left(
0,
\frac{1+
\frac{ \epsilon^2}{2} \frac{{\cQ}^2 + t}{\cQ^2 + \Bx t}}{ \sqrt{1+\epsilon^2}} \cos\phi
\pm i \sin\phi,
\mp i \cos\phi
+
\frac{1+\frac{\epsilon^2}{2} \frac{{\cQ}^2 + t}{\cQ^2 + \Bx t}}{\sqrt{1+\epsilon^2}}
\sin\phi,
\frac{-\epsilon \cQ \widetilde{K}/\sqrt{1+\epsilon^2}}{\cQ^2+\Bx t}
\right)
\, , \nonumber\\
\end{eqnarray}
which are defined up to an overall phase factor. The kinematical factor entering the
last component of $q_{2}$ reads
\begin{eqnarray}
{\widetilde K}
=
\sqrt{t_{\rm min} - t} \sqrt{(1 - \Bx)\sqrt{1 + \epsilon ^2}
+
\frac{(t_{\rm min} - t) \left(\epsilon^2 + 4 (1 - \Bx) \Bx \right)}{4 \cQ^2 }}
\, ,
\end{eqnarray}
and is related in an obvious manner to $K$ of Eq.\ \re{Kfactor} via
\begin{eqnarray}
K = \sqrt{1-y + \frac{\epsilon ^2}{4} y^2}\frac{{\widetilde K}}{\cQ}\,.
\end{eqnarray}
The photon polarization vectors obey the following completeness relations
\begin{eqnarray}
\label{Rel-ComHel-1}
\sum_{h=-,+} \varepsilon_1^\mu(h) \varepsilon_1^{\nu \ast}(h)
\!\!\!&-&\!\!\!
\varepsilon_1^\mu(0) \varepsilon_1^{\nu}(0) = - g^{\mu \nu} + \frac{q_1^\mu q_1^\nu}{q_1^2}
\, , \\
\label{Rel-ComHel-2}
\sum_{h=-,+} \varepsilon_2^\mu(h) \varepsilon_2^{\nu \ast}(h)
\!\!\!&=&\!\!\!
- g^{\mu \nu}
+
\frac{q_2^\mu p_1^\nu + p_1^\mu  q_2^\nu }{p_1\cdot q_2}
-
\frac{\epsilon^2 \cQ^2 q_2^\mu q_2^\nu }{( \cQ^2 + \Bx t)^2} \, ,
\end{eqnarray}
which are used in the main text to reduce the cross section to the product of helicity amplitudes.

\section{Fourier harmonics in leptonic tensor}
\label{FourierHarmonics}

Let us present explicit expressions for the Fourier coefficients entering the leptonic part
of the interference term \re{Def-Sqa-Int-Hel}. For the transverse-transverse harmonics we
found
\begin{eqnarray}
C_{++} (n \!\!\!&=&\!\!\! 0)
=
-
\frac{4(2 - y) \left(1+\sqrt{1+\epsilon^2}\right)}{(1+\epsilon^2)^{2}}
\Bigg\{
\frac{{\widetilde K}^2}{\cQ^2} \frac{(2-y)^2 }{\sqrt{1+\epsilon^2}}
\\
&+&\!\!\!
\frac{t}{\cQ^2} \left(1-y-\frac{\epsilon^2}{4} y^2 \right)(2-\Bx)
\Bigg(
1
+
\frac{2\Bx \left(2-\Bx + \frac{\sqrt{1+\epsilon^2}-1}{2}+ \frac{\epsilon^2}{2\Bx}\right)\frac{t}{\cQ^2}
+
\epsilon^2}{(2-\Bx)(1+\sqrt{1+\epsilon^2})}
\Bigg)
\Bigg\}
\, ,
\nonumber\\
C_{++} (n \!\!\!&=&\!\!\! 1)
=
\frac{-16 K\left(1-y-\frac{\epsilon^2}{4} y^2 \right)}{(1+\epsilon^2)^{5/2}}
\left\{
\left(
1+ (1-\Bx)  \frac{\sqrt{\epsilon ^2+1}-1}{2 \Bx} + \frac{\epsilon ^2}{4 \Bx}
\right)
\frac{\Bx t}{\cQ^2} - \frac{3 \epsilon ^2}{4}
\right\}
\nonumber\\
&-&\!\!\!\!\!
4 K \left(2-2 y+y^2+ \frac{\epsilon^2}{2}y^2\right)
\frac{1+\sqrt{1+\epsilon ^2}-\epsilon ^2}{(1+\epsilon^2)^{5/2}}
\Bigg\{\!
1 - (1-3\Bx) \frac{t}{{\cal Q}^2}
\nonumber\\
&&\qquad\qquad\qquad\qquad\qquad\qquad\qquad\qquad\qquad\qquad
+
\frac{1 - \sqrt{1 + \epsilon^2} + 3 \epsilon^2}{1 + \sqrt{1 + \epsilon^2} - \epsilon^2} \frac{\Bx t}{\cQ^2}
\Bigg\}
,\nonumber \\
C_{++} (n \!\!\!&=&\!\!\! 2)
= \frac{8(2-y)\left(1-y-\frac{\epsilon^2}{4} y^2 \right)}{(1+\epsilon^2)^{2}}
\Bigg\{ \frac{2\epsilon ^2}{1+\epsilon^2+\sqrt{1+\epsilon ^2}} \frac{{\widetilde K}^2}{\cQ^2}
\nonumber\\
&&\qquad\qquad\qquad\qquad\qquad\qquad\qquad\qquad
+
\frac{\Bx t\, t^{\prime}}{\cQ^4}
\left(
1-\Bx -\frac{\sqrt{1+\epsilon ^2}-1}{2} + \frac{\epsilon^2}{2\Bx}
\right)
\Bigg\}
\, ,
\nonumber \\
C_{++} (n \!\!\!&=&\!\!\! 3)
= -8 K \left(1-y-\frac{\epsilon^2}{4} y^2 \right)
\frac{\sqrt{1+\epsilon ^2}-1}{(1+\epsilon^2)^{5/2}}
\left\{
(1 - \Bx) \frac{t}{{\cal Q}^2}
+ \frac{\sqrt{1+\epsilon ^2}-1}{2} \left( 1 + \frac{t}{{\cal Q}^2} \right)
\right\}
\, ,
\nonumber\\
S_{++} (n \!\!\!&=&\!\!\! 1)
= -\frac{8 K  (2 - y)y}{1 + \epsilon^2}
\Bigg\{ 1+
\frac{1-\Bx+\frac{\sqrt{1+\epsilon^2}-1}{2}}{1+\epsilon^2}\,\frac{t^{\prime}}{\cQ^2}
\Bigg\}
\, ,
\nonumber\\
S_{++} (n \!\!\!&=&\!\!\! 2)
\nonumber\\
&=&\!\!\!
\frac{4\left(1-y-\frac{\epsilon^2}{4} y^2\right)y}{(1+\epsilon^2)^{3/2}}
\left(1+\sqrt{1+\epsilon^2} -2 \Bx\right)
\frac{t^{\prime}}{\cQ^2}
\Bigg\{\frac{\epsilon^2- \Bx (\sqrt{1+\epsilon^2}-1)}{1+\sqrt{\epsilon^2+1} -2 \Bx}
-
\frac{2 \Bx+\epsilon^2}{2\sqrt{1+\epsilon^2}}\,\frac{t^{\prime}}{\cQ^2}
\Bigg\}
, \nonumber
\end{eqnarray}
while we got for the longitudinal-transverse ones,
\begin{eqnarray}
C_{0 +} (n \!\!\!&=&\!\!\! 0)
=
\frac{12 \sqrt{2} K (2-y) \sqrt{1-y-\frac{\epsilon^2}{4} y^2} }{\left(1+\epsilon^2\right)^{5/2}}
\left\{
\epsilon^2+ \frac{2-6 \Bx-\epsilon^2}{3} \frac{t}{\cQ^2}
\right\}
\, ,
\\
C_{0 +} (n \!\!\!&=&\!\!\! 1)
=
\frac{8\sqrt{2} \sqrt{1-y-\frac{\epsilon^2}{4} y^2 }}{\left(1+\epsilon ^2\right)^2}
\Bigg\{
(2-y)^2 \frac{t^\prime}{\cQ^2}
\Bigg(
1 - \Bx +  \frac{(1-\Bx) \Bx+ \frac{\epsilon^2}{4}}{\sqrt{1+\epsilon^2}} \frac{t^\prime}{\cQ^2}
\Bigg)
\nonumber\\
&&\qquad
+\frac{1 - y - \frac{\epsilon^2}{4} y^2 }{\sqrt{1+\epsilon^2}} \left(1-(1-2 \Bx) \frac{t}{\cQ^2}\right)
\left(\epsilon ^2- 2 \left(1+\frac{\epsilon ^2}{2 \Bx}\right) \frac{\Bx t }{\cQ^2} \right)
\Bigg\}
\, ,
\nonumber\\
C_{0 +} (n \!\!\!&=&\!\!\! 2)
=
-\frac{8 \sqrt{2} K (2-y)\sqrt{1-y-\frac{\epsilon^2}{4} y^2} }{\left(1+\epsilon^2\right)^{5/2}}
\left(1+\frac{\epsilon^2}{2}\right)
\left\{
1 + \frac{1 +\frac{\epsilon^2}{2\Bx}}{1+\frac{\epsilon^2}{2}} \frac{\Bx t}{\cQ^2 }
\right\}
\, , \nonumber\\
S_{0 +} (n \!\!\!&=&\!\!\! 1)
=
-\frac{8 \sqrt{2} (2-y) y \sqrt{1-y-\frac{\epsilon^2}{4} y^2}}{
\left(1+\epsilon^2\right)^2} \frac{{\widetilde{K}^2}}{\cQ^2}
\, , \nonumber\\
S_{0 +} (n \!\!\!&=&\!\!\! 2)
=
-\frac{8 \sqrt{2} K y \sqrt{1-y-\frac{\epsilon^2}{4} y^2}}{\left(1+\epsilon^2\right)^2}
\left(1+\frac{\epsilon^2}{2}\right)
\left\{
1 + \frac{1 +\frac{\epsilon^2}{2\Bx}}{1+\frac{\epsilon^2}{2}} \frac{\Bx t}{\cQ^2 }
\right\}
\, .
\nonumber
\end{eqnarray}
Finally, the helicity-flip transverse-transverse coefficients are
\begin{eqnarray}
C_{-+} (n \!\!\!&=&\!\!\! 0)
= \frac{8 (2-y)}{\left(1+\epsilon^2\right)^{3/2}}
\Bigg\{
(2-y)^2 \frac{\sqrt{1+\epsilon ^2}-1}{2(1+\epsilon^2)} \frac{\widetilde{K}^2}{\cQ^2}
\\
&&\quad
+\frac{1-y-\frac{\epsilon^2}{4} y^2}{\sqrt{1+\epsilon^2}}
\Bigg(
1 -\Bx-\frac{\sqrt{1+\epsilon^2}-1}{2}+ \frac{\epsilon^2}{2\Bx}
\Bigg) \frac{\Bx t\, t^\prime}{\cQ^4}
\Bigg\}
\, ,
\nonumber\\
C_{-+} (n \!\!\!&=&\!\!\! 1)
=
\frac{8 K}{\left(1+\epsilon^2\right)^{3/2}}
\Bigg\{ (2-y)^2 \frac{2-\sqrt{1+\epsilon ^2}}{1+\epsilon ^2}
\Bigg(
\frac{\sqrt{1+\epsilon ^2}-1+\epsilon^2}{2 \left(2-\sqrt{1+\epsilon ^2}\right)}
\left(1-\frac{t}{\cQ^2}\right)-
\frac{\Bx t}{\cQ^2}
\Bigg)
\nonumber\\
&+&\!\!\!
2 \frac{1-y-\frac{\epsilon^2}{4} y^2}{\sqrt{1+\epsilon^2}}
\Bigg(
\frac{1-\sqrt{1+\epsilon^2}+\frac{\epsilon^2}{2}}{2 \sqrt{1+\epsilon^2}}
+
\frac{t}{\cQ^2}
\left( 1-\frac{3 \Bx}{2}+\frac{\Bx+\frac{\epsilon ^2}{2}}{2 \sqrt{1+\epsilon^2}}\right)
\Bigg)
\Bigg\}
\, ,
\nonumber\\
C_{-+} (n \!\!\!&=&\!\!\! 2)
=
4(2-y) \left(1 - y - \frac{\epsilon^2}{4} y^2\right)
\frac{1 + \sqrt{1+\epsilon ^2}}{\left(1+\epsilon^2\right)^{5/2}}
\Bigg\{
(2 - 3\Bx) \frac{t}{\cQ^2}
\nonumber\\
&+&\!\!\!
\left(1-2 \Bx+  \frac{2 (1-\Bx)}{1+\sqrt{1+\epsilon^2}}\right)\frac{\Bx t^2}{\cQ^4}
+
\Bigg(
1 + \frac{\sqrt{1+\epsilon^2} + \Bx + (1-\Bx)\frac{t}{\cQ^2}}{1+\sqrt{1+\epsilon^2}} \frac{t}{\cQ^2} \Bigg)
\epsilon^2
\Bigg\}
\, , \nonumber\\
C_{-+} (n \!\!\!&=&\!\!\! 3)
=
-8K \left(1-y-\frac{\epsilon^2}{4} y^2 \right)
\frac{1+ \sqrt{1+\epsilon ^2}+\frac{\epsilon^2}{2}}{\left(1+\epsilon^2\right)^{5/2}}
\Bigg\{
1+ \frac{1+\sqrt{1+\epsilon ^2}+\frac{\epsilon^2}{2 \Bx }}{1+\sqrt{1+\epsilon ^2}
+
\frac{\epsilon^2}{2}} \frac{\Bx t}{\cQ^2} \Bigg\}
\, ,
\nonumber\\
S_{-+} (n \!\!\!&=&\!\!\! 1)
=
-\frac{4 K (2-y) y}{\left(1+\epsilon^2\right)^2}
\Bigg\{
1 - \sqrt{1+\epsilon^2}+ 2 \epsilon^2 - 2 \left( 1 + \frac{\sqrt{1+\epsilon^2}-1}{2\Bx} \right) \frac{\Bx t}{\cQ^2}
\Bigg\}\,,
\nonumber\\
S_{-+} (n \!\!\!&=&\!\!\! 2)
= - 2 y \left(1-y-\frac{\epsilon^2}{4} y^2 \right)
\frac{1+\sqrt{1+\epsilon^2}}{\left(1+\epsilon^2\right)^2}
\Bigg(\epsilon^2 - 2 \Bigg( 1 +\frac{\epsilon^2}{2\Bx} \Bigg) \frac{\Bx t}{\cQ^2} \Bigg)
\nonumber\\
&&\qquad
\times\Bigg\{1+
\frac{\sqrt{1+\epsilon^2}-1+ 2\Bx}{1+\sqrt{1+\epsilon^2}}
\frac{t}{\cQ^2} \Bigg\}
\, . \nonumber
\end{eqnarray}

\end{document}